\documentclass[12pt,epsfig]{article}
\usepackage{graphicx}
\usepackage{amsmath,amssymb}
\usepackage[latin1]{inputenc}

\begin{document}
\begin{titlepage}

\title{Approximate solutions in General Relativity
 \\via deformation of embeddings}
 \author{Richard Kerner$^{\star}$ and Salvatore Vitale$^{\star}$}

\maketitle



\begin{abstract}

{\small A systematic study of deformations of four-dimensional Einsteinian space-times
embedded in a pseudo-Euclidean space $E^N$ of higher dimension is presented. Infinitesimal
deformations, seen as vector fields in $E^N$, can be divided in two parts, tangent
to the embedded hypersurface and orthogonal to it; only the second ones are relevant,
the tangent ones being equivalent to coordinate transformations in the embedded manifold.

The geometrical quantities can be then expressed in terms of embedding functions $z^A$
and their infinitesimal deformations $v^A \, z^A \rightarrow {\tilde{z}}^A = z^A + \varepsilon \, v^A$.
The deformations are called Einsteinian if they keep Einstein equations satisfied up to
a given order in $\varepsilon$. The system so obtained is then analyzed in particular
in the case of the Schwarzschild metric taken as the starting point, and some solutions of
the first-order deformation of Einstein's equations are found.

We discuss also second and third order deformations leading to wave-like solutions and to the departure
from spherical symmetry towards an axial one (the approximate Kerr solution)}
\end{abstract}
\footnote{ \small $^{\star}$
  Laboratoire de Physique Th\'eorique de la Mati\`ere Condens\'ee, \\\small
  Universit\'e Pierre-et-Marie-Curie - CNRS UMR 7600 \\\small
  Tour 22, 4-\`eme \'etage, Bo\^{i}te 142, \\\small
  4, Place Jussieu, 75005 Paris, France\\\small
  \small}
\end{titlepage}




\newcommand{\ud}{\mathrm{d}}
\newcommand{\go}{\mathop{g}^{{\rm o}}}

\newcommand{\gu}{\mathop{g}^{1}}

\newcommand{\gd}{\mathop{g}^{2}}

\newcommand{\gamo}{\mathop{\Gamma}^{{\rm o}}}

\newcommand{\gamu}{\mathop{\Gamma}^{1}}

\newcommand{\gamd}{\mathop{\Gamma}^{2}}

\newcommand{\Rio}{\mathop{R}^{{\rm o}}}

\newcommand{\Riu}{\mathop{R}^{1}}

\newcommand{\Rid}{\mathop{R}^{2}}
\newcommand{\Rit}{\mathop{R}^{3}}
\newcommand{\vo}{\mathop{v}^{{\rm o}}}

\newcommand{\vu}{\mathop{R}^{1}}

\newcommand{\vd}{\mathop{v}^{2}}
\newpage
\section{Introduction}

The study of the two-body problem in General Relativity must include possible emission
of gravitational waves. In this respect General Relativity runs into difficulties akin
to those of classical electromagnetism, where the full treatment of charged particle motion
plus the radiative field can be performed only via approximation techniques. Great progress was
achieved since the first papers by Ll. Bel and N. Deruelle, \cite{LlBel}, followed by the
throughout analysis of post-Newtonian and post-post-Newtonian approximations
including gravitational radiation. The most important contributions have been made
by Th. Damour, L. Blanchet, G. Shaeffer and many others, \cite{Damour}, \cite{Blanchet},
\cite{Schaeffer}.

Their approach was based on a particular choice of the initial approximation. In physical
situations where the gravitational field is not too strong and when the relative velocities
of bodies under consideration are very low as compared with the speed of light $c$,
the Newtonian theory provides us with exact (although not explicit) solutions. One takes
one of these solutions as a starting point, adding successive corrections resulting from
the inclusion of the relativistic effects: the finite propagation of gravitational field,
the curvature of space, and so forth. The corrections are consequently made to the trajectory,
to the law of motion, and to the gravitation potential (identified with the corrections
to the space-time metric tensor).

In a series of papers published a few years ago \cite{Geodevi2001}, \cite{Motions2000},
\cite{Epicycles2001}, \cite{EpiKerr2002}) an alternative method of determining
relativistic motion of test particles in a spherically symmetric gravitational
field has been proposed, without need to use the Newtonian limit of General Relativity.
The idea was to take as a starting point the very special explicit solution of geodesic equation
in Schwartzschild metric background: a test particle moving with constant speed
along the circle around a spherically symmetric mass. Let us denote this particular
trajectory by $x^{\mu} \, (s)$, with $s$ denoting the proper spacetime length of the curve.
Then one can consider a deviation from this worldline, ${\tilde x}^{\mu} \, (s)$,
which can be expanded in a power series of some small parameter $\varepsilon$:

\begin{eqnarray}
{\tilde x}^{\mu} \, (s) = {x}^{\mu} \, (s) + \varepsilon \, n^{\mu} \, (s) + \varepsilon^2 \,
b^{\mu} \, (s) + ...
\label{geodevi1}
\end{eqnarray}

The geodesic equation in Schwarzschild background can be also expanded in a series of
equations whose solutions in terms of unknown deviations $n^{\mu} \, (s), \, b^{\mu} \, (s)$,
etc., will provide us with successive approximations to the exact solution, up to any
order required. The parameter $\varepsilon$ is roughly proportional to the eccentricity
of the new trajectory. Already the first approximation, linear in $\varepsilon$, predicts
the perihelion advance for orbits with very small eccentricity.

The shortcoming of this method, which was entirely focused on the trajectories, was
the total lack of any variation of the gravitational field, i.e. the Schwarzschild metric
which was maintained invariable for all orders of geodesic deviation. This fact reduced the validity
of the method only to the case of test particles with mass $m$ negligible when compared with
the mass $M$ of the central body appearing in the Schwarzschild background metric.
More precisely, the successive approximations of planet's trajectory remain valid as
long as the dimensionless parameter $(m/M)$ can be considered as negligibly small,
i.e. $(m/M) <<1$. When the mass $m$ is no more a negligible quantity, its presence
must inevitably alter the geometry of the initial Schwarzschild metric, and its
influence can be therefore represented by a power series in the small parameter $(m/M)$.

In this article we shall describe the departure from the initial
Schwarzschild metric in terms of the embedding functions. Embeddings
of the exterior Schwarzschild geometry in pseudo-Euclidean flat
spaces are known since a long time (\cite{Kasner1921},
\cite{Fronsdal1959}, \cite{Rosen1965}), and once such an embedding
is given, all intrinsic geometric quantities of the embedded
manifold can be expressed in terms of derivatives of the embedding
functions which depend on four ``internal'' parameters which are the
space-time coordinates. A general analysis of deformations of the
embedded Einstein spaces was given in \cite{Kerner1978}; nevertheless,
only the theoretical setup was considered, without any concrete solution
describing Ricci-flat deformations of known exact Einstein spaces,
and in the first place, Schwarzschild and Kerr metrics.

The present article is intended to explore not only the first linear
approximation, but also the effects of second and third order in
the expansion of deformations in powers of small parameter $\epsilon$,
including the corrections describing gravitational waves. The departure
from spherical symmetry (Schwarzschild's metric) towards axial symmetry
(Kerr's metric) as Einsteinian (Ricci-flat) deformation
of the corresponding embedded manifold is also discussed.

\section{Isometric embeddings and their properties}

\subsection{The embedding functions and the induced metric}

Consider the embedding of a four-dimensional Riemannian space parametrized by local
coordinates (denoted by $x^{\mu}$, $\mu, \nu = 0,1,2,3$ as usual) in a pseudo-Euclidean
space $E^N$ of dimension $N$. The dimension $N$, yet unspecified, depends on the topology
of the Riemannian space under consideration, and may be quite high, as acknowledged in \cite{Rosen1965}.
Locally, any n-dimensional Riemannian manifold can be embedded in a (pseudo)-Euclidean space
of dimension $N = n(n+1)/2$. Here we are interested in {\it global} embeddings, which may require
a relatively low dimension of the ``host'' space if the Riemannian space to be embedded possesses
some particular symmetry. For example, the de Sitter space can be embedded globally in
a five-dimensional pseudo-Euclidean space with signature $(+----)$, and both exterior
and interior Schwarzschild solutions can be embedded globally in a six-dimensional
$E^N$ with signatures $(++----)$ or $(+-----)$.
Consider a global embedding of a Riemannian space $V_4$ given by the following set of
{\it embedding functions} $z^A$:

\begin{eqnarray}
z^A = z^A\, (x^{\mu}), \,\;\quad {\rm with}\quad\;\; \begin{array}{ccl}
 A,B,... &=& 1,2,...N, \\ \mu, \nu &=& 0,1,2,3.\end{array}
\end{eqnarray}

The metric tensor of $V_4$ is the {\it induced metric} defined as

\begin{eqnarray}
{\go}_{\mu \nu} =\eta_{AB} \, \partial_{\mu} z^A \, \partial_{\nu} z^B
\label{godef1}
\end{eqnarray}

The inverse metric tensor $\displaystyle{ {\go}\,^{\mu \nu}}$ cannot be obtained
directly from the embedding functions, but should be computed from the covariant components
as their inverse matrix. From now on we use the superscript notation in order to make difference
between the ``basic'' induced metric $\displaystyle{{\go}\,_{\mu \nu}}$ which will be considered
as a background, and its infinitesimal deformations expanded in terms of an infinitesimal
parameter $\varepsilon$ as follows:

\begin{eqnarray}
g_{\mu \nu} = {\go}_{\mu \nu} + \varepsilon \, {\gu}_{\mu \nu} + \varepsilon^2 \, {\gd}_{\mu \nu} + ...
\label{gexpansion1}
\end{eqnarray}
induced by the following deformation of the initial embedding functions:
\begin{eqnarray}
z^A \, (x^\mu) \rightarrow z^A \, (x^\mu) + \varepsilon \, {\mathop{v}^{1}}\,^A  \, (x^\mu)
+ \varepsilon^2  \,  {\mathop{v}^{2}}\,^A \, (x^\mu) + ...
\label{zetexpansion1}
\end{eqnarray}


When seen from the ambient pseudo-Euclidean space, the new embedded manifold ${\tilde V}_4$
is the result of an infinitesimal deformation of the initial manifold $V_4$ induced
by a vector field in $E^N_{(p,q)}$. It is quite obvious that on the embedded manifold
such a field can be decomposed into its normal part (in the sense of the pseudo-Euclidean
metric) and a part tangent to $V_4$. This last part induces an internal diffeomorphism
of $V_4$ and can be always implemented as a local coordinate transformation. Such
deformations do not have any physical meaning, but it is not always necessary to consider
exclusively the deformations orthogonal to the embedded $V_4$; sometimes a deformation
having non-vanishing both parallel and orthogonal parts can have less non-zero components
in the ambient space $E^N_{(p,q)}$ than its part orthogonal to the embedded $V_4$ manifold.

\subsection{Expressions for connection and curvature}

Our first aim is to express all important geometrical quantities e.g. the connection coefficients
and the curvature tensor, in terms of embedding functions $z^A$ and their partial derivatives.
Let us start with Christoffel connection
\begin{eqnarray}
 {{\gamo}}\, ^{\lambda}_{\mu \nu}  = \frac{1}{2} \, {\displaystyle{ {\go}\,^{\lambda \rho}}}
\biggl(\partial_{\mu} {\displaystyle{{\go}\,_{\nu \rho}}} +   \partial_{\nu} {\displaystyle{{\go}\,_{\mu \rho }}}
- \partial_{\rho} {\displaystyle{{\go}\,_{\mu \nu}}} \biggr).
\label{christoffel1}
\end{eqnarray}
>From the definition of ${\displaystyle{{\go}\,_{\mu \nu}}}$ (\ref{godef1}) we have the expression
for its partial derivatives:
\begin{eqnarray}
\partial_{\lambda} \, {\displaystyle{{\go}\,_{\mu \nu}}} =  \eta_{AB}\,
\biggl(\partial^2_{\lambda \mu} \, z^A \, \partial_{\nu} \, z^B
+ \partial_{\mu} \, z^A \, \partial^2_{\lambda \nu} \, z^B \biggr).
\label{dergo1}
\end{eqnarray}
When substituted into the definition (\ref{christoffel1}) it gives
\begin{eqnarray}
{{\gamo}}\, ^{\lambda}_{\mu \nu}  = \eta_{AB} \, {\displaystyle{ {\go}\,^{\lambda \rho}}} \,
\partial_{\rho} \, z^A \, \partial^2_{\mu \nu} \, z^B
\label{defgamo2}
\end{eqnarray}
Consider now the second covariant derivative of $z^B$:
\begin{eqnarray}
\nabla_{\mu} \, \nabla_{\nu} \, z^B = \partial^2_{\mu \nu} z^B -
{{\gamo}}\, ^{\lambda}_{\mu \nu} \, \partial_{\lambda} z^B .
\label{partialz1}
\end{eqnarray}
Therefore, we have
\begin{eqnarray}
\partial^2_{\mu \nu} z^B = \nabla_{\mu} \, \nabla_{\nu} \, z^B
+ {{\gamo}}\, ^{\lambda}_{\mu \nu} \, \partial_{\lambda} z^B ,
\label{defgamo3}
\end{eqnarray}
and of course, $\partial_{\mu} \, z^B = \nabla_{\mu} \, z^B$.
Substituting (\ref{defgamo3}) into (\ref{defgamo2}) we obtain the following identity:
\begin{eqnarray}
{{\gamo}}\, ^{\lambda}_{\mu \nu} = \eta_{AB} \, {\displaystyle{ {\go}\,^{\lambda \rho}}} \,
\nabla_{\rho} \, z^A \, \nabla_{\mu} \nabla_{\nu} \, z^B
+ \eta_{AB} \, {\displaystyle{ {\go}\,^{\lambda \rho}}} \,
{{\gamo}}\, ^{\sigma}_{\mu \nu} \, \nabla_{\sigma} \, z^A \, \nabla_{\rho} \, z^B.
\label{christoffel3}
\end{eqnarray}
But in the last term we note that
\begin{eqnarray}
{\displaystyle{ {\go}\,^{\lambda \rho}}} \,  \eta_{AB} \,
\nabla_{\sigma} \, z^A \, \nabla_{\rho} \, z^B = {\displaystyle{ {\go}\,^{\lambda \rho}}} \,
{\displaystyle{ {\go}\,_{\sigma \rho}}} = \delta^{\lambda}_{\sigma},
\label{gamidentity}
\end{eqnarray}
reducing (\ref{christoffel3}) to
\begin{eqnarray}
{{\gamo}}\, ^{\lambda}_{\mu \nu} = \eta_{AB} \, {\displaystyle{ {\go}\,^{\lambda \rho}}}
\nabla_{\rho} \, z^A \, \nabla_{\mu} \nabla_{\nu} \, z^B + {{\gamo}}\, ^{\lambda}_{\mu \nu},
\label{identity4}
\end{eqnarray}
which shows clearly that
\begin{eqnarray}
\eta_{AB} \, \nabla_{\rho} \, z^A \, \nabla_{\mu} \nabla_{\nu} \, z^B = 0
\label{identity5}
\end{eqnarray}
which may be considered as an alternative (although implicit) definition of Christoffel symbols,
and could be also derived as a direct consequence of the fact that $\nabla_{\mu} \, g_{\lambda \rho} = 0$.

Using this result, let us form the following combination of covariant derivatives which
vanishes identically:
\begin{eqnarray*}
\eta_{AB} \, \, \biggl[ \nabla_{\mu} ( \nabla_{\rho} \, z^A \, \nabla_{\nu} \nabla_{\sigma} \, z^B )
- \nabla_{\nu} ( \nabla_{\rho} \, z^A \, \nabla_{\mu} \nabla_{\sigma} \, z^B ) \biggr] = 0
\label{GaussCodazzi1}
\end{eqnarray*}
Applying the derivation and using the Leibniz rule we get:
$$\eta_{AB} \, \, \biggl[ \nabla_{\mu}  \nabla_{\rho} \, z^A \, \nabla_{\nu} \nabla_{\sigma} \, z^B
- \nabla_{\nu} \nabla_{\rho} \, z^A \, \nabla_{\mu} \nabla_{\sigma} \, z^B ) \biggr] +$$
\begin{eqnarray}
+\, \eta_{AB} \, \nabla_{\rho} \, z^A \, \biggl[ \nabla_{\mu}  \nabla_{\nu} \nabla_{\sigma} \, z^B  -
\nabla_{\nu} \nabla_{\mu} \nabla_{\sigma} \, z^B  \biggr] = 0 .
\label{GaussCodazzi2}
\end{eqnarray}
Recalling that
\begin{eqnarray}
\biggl[\nabla_{\mu} \, \nabla_{\nu}  - \nabla_{\nu} \, \nabla_{\mu} \biggr] \, \nabla_{\rho} \, z^B =
{\Rio} \, ^{\, \ \ \, \lambda}_{\mu \nu \, \ \  \rho} \, \nabla_{\lambda} \, z^B ,
\label{GausCodazzi3}
\end{eqnarray}
so that we can write
\begin{eqnarray}
{\Rio}\, _{\mu \nu \, \lambda  \rho} = - \eta_{AB} \, \biggl[ \nabla_{\mu} \nabla_{\lambda} \, z^A
\nabla_{\nu} \nabla_{\rho} \, z^B -\nabla_{\nu} \nabla_{\lambda} \, z^A
\nabla_{\mu} \nabla_{\rho} \, z^B \biggr]
\label{GaussCodazzi4}
\end{eqnarray}
which is the well known Gauss-Codazzi equation.

The definition of the Riemann tensor by means of derivatives of the embedding functions given by formula
(\ref{GaussCodazzi4}) looks very compact, but is in fact highly non-linear and quite complicated. This is so
because it contains many Christoffel symbols involved in the second covariant derivatives, which contain in turn
the contravariant metric tensor $\displaystyle{{\go}} \,^{\mu \nu}$. The components of the contravariant
metric tensor are obtained as rational expressions in third and fourth powers of $\nabla_{\mu}\,z^A$.
Nevertheless, the most important point here is that the Riemann tensor depends only on first and second derivatives
of embedding functions, so that the Einstein equations expressed in terms of the embedding functions will lead
to second-order partial differential equations.

The expressions derived in this section will be very useful in the development of a power series
expansion of infinitesimally deformed embedding.

\section{Infinitesimal deformations of embeddings}

\subsection{General setting}
Let us consider an isometric embedding of an Einsteinian manifold ${\displaystyle{\mathop{V}^{\rm o}}}_4$
in a pseudo-Euclidean space $E^N_{p,q}$ with signature $(p +, q -)$, with $p+q = N$:
\begin{eqnarray}
\begin{array}{ccl}
 {\displaystyle{\mathop{V}^{\rm o}}}_4   &\rightarrow&   E^N_{p,q} \\   z^A &=& z^A\, (x^{\mu})\end{array}
\;\; {\rm with}\;\;  \begin{array}{ccl}
 A,B,... &=& 1,2,...N, \\ \mu, \nu &=& 0,1,2,3.\end{array}
\label{embedzero}
\end{eqnarray}
Consider now an infinitesimal deformation of the embedding defined by a converging series of terms
proportional to the consecutive powers of a small parameter $\varepsilon$. The deformed embedding
defines an Einsteinian space ${\tilde{V}}_4$;
\begin{eqnarray*}
z^A \, (x^\mu) \rightarrow {\tilde{z}}^A \, (x^{\mu}) = z^A \,(x^\mu) + \varepsilon \,
v\,^A  \, (x^\mu)
+ \varepsilon^2  \,  w^A \, (x^\mu) + \ldots
\end{eqnarray*}
The induced metric on ${\tilde{V}}_4$ can also be developed in series of powers of $\varepsilon$:
\begin{eqnarray}
{\tilde{g}}_{\mu \nu} &=& {\displaystyle{\go} \, _{\mu \nu}} + \varepsilon  {\displaystyle{\gu} \, _{\mu \nu}}
+ \varepsilon^2  {\displaystyle{\gd} \, _{\mu \nu}} + ...\nonumber \\
&=& \eta_{AB}  \Biggl[\partial_{\mu} z^A \partial_{\nu} z^B +  \varepsilon
\biggl( \partial_{\mu} z^A \partial_{\nu} v^B +
\partial_{\mu} v^A \partial_{\nu} z^B \biggr) + \nonumber\\
&+& \varepsilon^2 \biggl(
\partial_{\mu} v^A \partial_{\nu} v^B +\partial_{\mu} z^A \partial_{\nu} w^B +
\partial_{\mu} w^A \partial_{\nu} z^B  \biggr) \Biggr]
\label{defmetric1}
\end{eqnarray}

Among all possible infinitesimal deformations of the embedding functions $z^A \, (x^{\mu}) +
\varepsilon \, v^A \, (x^{\mu})$
there is a large class of functions $v^A \, (x^{\mu})$ which will not alter the intrinsic
geometry of the embedded manifold. Infinitesimal translations $v^A = Const.$ obviously do not
change the internal metric $ {\displaystyle{\go} \, _{\mu \nu}} = \eta_{AB} \, \partial_{\mu} z^A
\partial_{\nu} z^B$. Also the generalized Lorentz transformations of the pseudo-Euclidean space
$E_{(p,q)}^N$ keep the internal metric unchanged. Indeed, if we set
\begin{eqnarray}
z^A \rightarrow {\tilde{z}}^A = z^A + \varepsilon \, \Lambda^A_{\, \, \, B} \, z^B,
\label{LorentzN}
\end{eqnarray}
with $\Lambda^A_{\, \, \, B}$ constant matrix. Then the first-order deformed metric is:

\begin{eqnarray*}
{\tilde{g}}_{\mu \nu} &=& {\displaystyle{\go} \, _{\mu \nu}} +
\varepsilon  {\displaystyle{\gu} \, _{\mu \nu}} + \ldots
=\nonumber  \\
 &=&\eta_{AB}  \Biggl[\partial_{\mu} z^A \partial_{\nu} z^B +
\varepsilon \, \biggl( \partial_{\mu} z^A  {\Lambda}^B_{\, \, C} \partial_{\nu} z^C +
{\Lambda}^A_{\, \,  C} \partial_{\nu} z^C \partial_{\mu} z^B \biggr) \Biggr]=\nonumber \\
&=& {\displaystyle{\go} \, _{\mu \nu}} + \varepsilon \, \biggl[ \eta_{AB} {\Lambda}^B_{\, \, C}
+ \eta_{CB} {\Lambda}^C_{\, \, A} \biggr] \partial_{\mu} z^A \partial_{\nu} z^B\qquad \qquad ,
\label{defmetricnull}
\end{eqnarray*}

Then the first-order correction vanishes if the matrices $\Lambda^A_{\, \, B}$ satisfy the
identity
$$\eta_{AB} \, {\Lambda}^B_{\, \, C}
+ \eta_{CB} {\Lambda}^C_{\, \, A} = 0$$

which defines the infinitesimal rigid rotations (Lorentz transformations) of the pseudo-Euclidean
space $E_{(p,q)}^N$.

The geometric character of our approach enables us to eliminate unphysical degrees of freedom
using simple geometrical arguments. Remember that in the traditional approach leading to
linearized equations for gravitational fields the starting point is the following development
of the metric tensor:
\begin{eqnarray}
g_{\mu \nu} = {\displaystyle{\go} \, _{\mu \nu}} + \varepsilon \, h_{\mu \nu},
\label{hamunu}
\end{eqnarray}
thus introducing {\it ten} components of $h_{\mu \nu}$ as dynamical fields. We know however that
most of them do not represent real dynamical degrees of freedom due to the gauge invariance.
The metric tensor itself does not correspond to any directly measurable quantity. In fact,
its components may be changed by a gauge transformation without changing the components of
the Riemann tensor which is the source of measurable gravitational effects. In particular,
the gauge transformation
\begin{eqnarray}
g_{\mu \nu} \rightarrow {\tilde{g}}_{\mu \nu} = g_{\mu \nu} + \nabla_{\mu} \, \xi_{\nu}
+ \nabla_{\nu} \, \xi_{\mu}
\label{gauge1}
\end{eqnarray}
does not alter the Riemann tensor so that both $\tilde{g}_{\mu \nu}$ and $g_{\mu \nu}$
describe the same gravitational field.

The arbitrary vector field $\xi^{\mu}$ generating gauge transformation (\ref{gauge1}) represents
four degrees of freedom which are redundant in $g_{\mu \nu}$; this is why in the linearized
Einstein equations one may impose four gauge conditions e.g.
\begin{eqnarray}
\nabla_{\mu} \, h^{\mu \nu} = 0.
\label{gauge2}
\end{eqnarray}

The unphysical degrees of freedom can be easily eliminated from the embedding deformation
functions $v^A(x^{\mu})$ if we note that any vector field in the embedding space $E^N$
that is {\it tangent} to the embedded Riemannian space $V_4$ describes nothing else but
a diffeomorphism of $V_4$, in other words a coordinate change, which has no influence on
any physical or geometrical quantities.

Vector fields tangent to the four-dimensional embedded manifold $V_4$ can be decomposed
along four arbitrarily chosen independent smooth vector fields in $E^N$ tangent
to $V_4$. On the other hand, vector fields {\it transversal} to the embedded hypersurface
$V_4$ must satisfy the following obvious {\it orthogonality conditions}:
\begin{eqnarray}
\eta_{AB} \, \partial_{\mu} \, z^A \, v^B = \eta_{AB} \, v^A \, \nabla_{\mu} \, z^B = 0.
\label{transversality1}
\end{eqnarray}
For any value of $A$ the four partial derivatives (let us remind that $\nabla_{\mu} \, z^A
= \partial_{\mu} \, z^A)$ span a basis of four vector fields in $E^N$ tangent to the
submanifold $V_4$; therefore any vector $v^B$ satisfying the orthogonality condition
(\ref{transversality1}) is transversal to $V_4$ (as seen in $E^N$).

The orthogonality condition (\ref{transversality1}) imposes four independent equations,
which reduce the number of independent deformation functions $v^A$ to $N-4$. This means
that general non-redundant deformations can be decomposed along $N-4$ independent fields
$X^A_{(k)}, \, \ \ k=1,2,...,N-4$:
\begin{eqnarray}
v^A (x^{\mu}) = {\displaystyle{\sum_{k=1}^{N-4}}} \, v^{k} (x^{\nu})) \, X^A_{(k)} \, (x^{\lambda}).
\label{decomp}
\end{eqnarray}
The basic fields $X^A_{(k)} \, (x^{\lambda})$ can be chosen at will provided they induce
a non-singular global vector field on $V_4$, while the relevant degrees of freedom are contained
in $N-4$ functions $v^k(x^{\lambda})$. To take an example, the de Sitter space can be globally
embedded in a five-dimensional pseudo-Euclidean space $E^5_{1,4}$ with signature $(+----)$;
therefore its global deformations can be described by a single function $v(x^{\lambda})$
(see \cite{Kerner1978}).

One may ask the following question: if the deformation destroys the initial symmetry of the
embedded manifold so that the deformed manifold cannot be embedded in the initially sufficient
$N$-dimensional pseudo-Euclidean space but needs a flat embedding space of higher dimension ?
It is known that global embeddings of the Kerr metric need more than six flat dimensions
sufficient for the embedding of the exterior (or interior) Schwarzschild solution (see
\cite{Carter1967}, \cite{Kuzeev1981}, \cite{Romero1996}, \cite{Romero1997}), although Schwarzschild's
metric can be obtained from Kerr's metric as
a limit when the Kerr parameter $a$ (the angular momentum) tends to zero.

The answer is that as long as we investigate only the first-order corrections to geometry,
we should not worry about this issue for the following two reasons: first, when a global
embedding is given, its infinitesimal deformations cannot lead to a global modification of
the embedding; second, if the bigger embedding space was introduced, say $E^{N+m}$, it would contain
the initial embedding space $E^N$ as its linear subspace, so that
$$E^{N+m} = E^N \oplus E^m, $$
and its pseudo-Euclidean metric could be represented as a blockwise reducible matrix
\begin{eqnarray}
\eta_{\alpha \beta} =
\begin{pmatrix} \eta_{AB} & 0 \cr
0 & \eta_{ij}
\end{pmatrix},
\label{bigmetric}
\end{eqnarray}
with $ \, A,B,...=1,2,...N, \, \, i,j,...= 1,2,..m, \, \, \alpha, \beta,...
= 1,2,....,N+m.$
Accordingly, any deformation of the initial embedding can be decomposed in two parts,
one contained in the initial embedding space $E^N$ and another one in the complementary
subspace $E^m$:
\begin{eqnarray}
v^{\alpha} = [\, v^A, \, v^m \, ].
\label{vdecomp}
\end{eqnarray}
But the initial embedding functions had their components entirely in the first subspace $E^N$,
$z^{\alpha} = [ \, z^A, \, 0 \, ]$, therefore the deformed embedding functions can be written as
\begin{eqnarray}
{\tilde{z}}^{\alpha} = [\, z^A + \varepsilon \, v^A, \, \varepsilon \, v^k \, ],
\label{zdecomp}
\end{eqnarray}
so that the induced metric of the deformed embedding will be
\begin{eqnarray}
g_{\mu \nu} &=& {\displaystyle{\go} \, _{\mu \nu}} + \varepsilon  {\displaystyle{\gu} \, _{\mu \nu}}
+ \varepsilon^2  {\displaystyle{\gd} \, _{\mu \nu}} + ...\nonumber\\
 &=& \eta_{AB}  \biggl(\partial_{\mu} z^A \partial_{\nu} z^B \biggr) +
\varepsilon \, \eta_{AB} \,  \biggl( \partial_{\mu} z^A \partial_{\nu} v^B +
\partial_{\mu} v^A \partial_{\nu} z^B \biggr)\\
&+& \varepsilon^2 \, \Biggl[ \eta_{AB} \biggl(
\partial_{\mu} v^A \partial_{\nu} v^B +\partial_{\mu} z^A \partial_{\nu} w^B +
\partial_{\mu} w^A \partial_{\nu} z^B  \biggr)+ \eta_{ij} \, \biggl(\partial_{\mu}\, v^i \partial_{\nu} \, v^j \, \biggr) \Biggr] ...\nonumber
\label{defmetric3}
\end{eqnarray}
>From this one can see that the deformations towards the extra dimensions do not
contribute to the first-order corrections of any geometrical quantities obtained from
the deformed embedding functions. This is why we shall not consider such deformations
while investigating at first only the terms linear in the infinitesimal parameter
$\varepsilon$.
Our principal aim now is to establish the explicit form of connection and curvature
components induced on the infinitesimally deformed embedding ${\tilde V}_4$. To this end
we must calculate the approximate expression of the contravariant metric tensor $g^{\mu \nu}$.
If the covariant metric is decomposed as
\begin{eqnarray}
g_{\mu \nu} = {\go}_{\mu \nu} + \varepsilon \, {\gu}_{\mu \nu} + \varepsilon^2 \, {\gd}_{\mu \nu} + ...
\label{gexpansion2}
\end{eqnarray}
then we have the following formulae defining the corresponding decomposition of $g^{\mu \nu}$:
\begin{eqnarray}
g^{\mu \nu} &=& {\go}\,^{\mu \nu} + \varepsilon \, {\gu}\,^{\mu \nu}
+ \varepsilon^2 \, {\gd}\,^{\mu \nu} + \ldots \nonumber\\
&=& g^{\mu \nu} - \varepsilon \, {\go}\,^{\mu \rho} \,  {\go}\,^{\mu \sigma} \,
{\gu}_{\rho \sigma}
- \varepsilon^2 \, \Biggl[ {\go}\,^{\mu \rho}\, {\go}\,^{\mu \sigma} \, {\gd}_{\rho \sigma} \Biggl]+ \varepsilon^2 \, \Biggl[ {\go}\,^{\mu \rho} \, {\go}\,^{\mu \sigma}\,  {\go}\,^{\lambda \kappa} \, \,
{\gu}_{\rho \lambda} \, {\gu}_{ \sigma \kappa} \Biggr].
\label{gcontrdecomp}
\end{eqnarray}

\subsection{The first order corrections to Einstein equations}

In what follows we shall keep only the first order terms linear in $\varepsilon$.

Let us start by computing the first (linear) correction to the components of the Christoffel connection,
which develops in Taylor series as
\begin{eqnarray}\label{Chrisexp}
\Gamma^\lambda_{\mu \nu} = {\displaystyle{\gamo}}\,^{\lambda}_{\mu \nu}
+ \varepsilon \, {\displaystyle{\gamu}}\,^{\lambda}_{\mu \nu}
+ {\displaystyle{\gamd}}\,^{\lambda}_{\mu \nu} + ...,
\end{eqnarray}
then by definition we have:
\begin{eqnarray}
{\displaystyle{\gamu}}\,^{\lambda}_{\mu \nu}  &=& \frac{1}{2} \, {\displaystyle{ {\go}\,^{\lambda \rho}}}
\biggl(\partial_{\mu} {\displaystyle{{\gu}\,_{\nu \rho}}} +   \partial_{\nu} {\displaystyle{{\gu}\,_{\mu \rho }}}- \partial_{\rho} {\displaystyle{{\gu}\,_{\mu \nu}}} \biggr) + \frac{1}{2} \, {\displaystyle{{\gu}\,^{\lambda \rho}}}\biggl(\partial_{\mu} {\displaystyle{{\go}\,_{\nu \rho}}} +\partial_{\nu} {\displaystyle{{\go}\,_{\mu \rho }}}- \partial_{\rho} {\displaystyle{{\go}\,_{\mu \nu}}} \biggr)\\
&=&  \frac{1}{2} \, {\displaystyle{\go}}\,^{\lambda \rho}
\biggl(\partial_{\mu} {\displaystyle{{\gu}\,_{\nu \rho}}} + \partial_{\nu} {\displaystyle{{\gu}\,_{\mu \rho }}}- \partial_{\rho} {\displaystyle{{\gu}\,_{\mu \nu}}} \biggr)\-  \frac{1}{2} \, {\displaystyle{ {\go}\,^{\lambda \sigma}}} \,{\displaystyle{ {\go}\,^{\rho \kappa}}}{\displaystyle{ {\gu}\,_{\sigma \kappa}}}\biggl(\partial_{\mu} {\displaystyle{{\go}\,_{\nu \rho}}} +   \partial_{\nu} {\displaystyle{{\go}\,_{\mu \rho }}}- \partial_{\rho} {\displaystyle{{\go}\,_{\mu \nu}}} \biggr).\nonumber
\label{gamofirst1}
\end{eqnarray}
One easily checks that
\begin{eqnarray}
\frac{1}{2} \, {\displaystyle{\go}}\,^{\lambda \rho}
\biggl(\partial_{\mu} {\displaystyle{{\gu}\,_{\nu \rho}}} + \partial_{\nu} {\displaystyle{{\gu}\,_{\mu \rho }}}
- \partial_{\rho} {\displaystyle{{\gu}\,_{\mu \nu}}} \biggr)=
 \eta_{AB} \, {\go}\,^{\lambda \rho} \, \biggl[ \, \partial_{\rho} z^A \, \partial^2_{\mu \nu} \, v^B
+ \partial_{\rho} v^A \, \partial^2_{\mu \nu} \, z^B \, \biggr],
\label{gamofirst2}
\end{eqnarray}
while the first term after some algebra gives
\begin{eqnarray}
 \frac{1}{2} \, {\displaystyle{ {\gu}\,^{\lambda \rho}}}
\biggl(\partial_{\mu} {\displaystyle{{\go}\,_{\nu \rho}}} +   \partial_{\nu} {\displaystyle{{\go}\,_{\mu \rho }}}
- \partial_{\rho} {\displaystyle{{\go}\,_{\mu \nu}}} \biggr) =
- \eta_{AB} \, {\go}\, ^{\lambda \rho} \biggl[ \,
\partial_{\rho} \, z^A {\gamo}\,^{\sigma}_{\mu \nu} \, \partial_{\sigma} \, v^B +
\partial_{\rho} \, v^A {\gamo}\,^{\sigma}_{\mu \nu} \, \partial_{\sigma} \, z^B \, \biggr].
\label{gamofirst3}
\end{eqnarray}
Combining together (\ref{gamofirst1}), (\ref{gamofirst2}) and (\ref{gamofirst3}) we find the final expression
\begin{eqnarray}
{\displaystyle{\gamu}}\,^{\lambda}_{\mu \nu} = \eta_{AB} \, {\go}\,^{\lambda \rho} \,
\biggl[ \, \nabla_{\rho} z^A \, \nabla_{\mu} \nabla_{\nu} \, v^B
+ \nabla_{\rho} v^A \, \nabla_{\mu} \nabla_{\nu} \, z^B \, \biggr]
\label{gamo1fin}
\end{eqnarray}
\indent
This expression has a tensorial character as it should be, because by definition both quantities
$$ \Gamma\,^{\lambda}_{\mu \nu} \, \ \ \, {\rm and} \, \ \ \,
{\tilde{\Gamma}}\,^{\lambda}_{\mu \nu}$$
transform as connection coefficients, therefore their difference must transform as a tensor, and
this is true for any term of the development into series of powers of $\varepsilon$.

\indent
The coefficients ${\displaystyle{\gamu}}\,^{\lambda}_{\mu \nu}$ will be useful for the derivation
of geodesic equations in the deformed space-time, but they are not necessary for the computation
of the first-order deformation of the Riemann tensor, which can be determined as follows.

Let us develop second covariant derivatives of the deformed embedding functions ${\tilde{z}}^A$ yields:

\begin{eqnarray}
{\tilde{ \nabla}}_{\mu} {\tilde{ \nabla}}_{\nu} \, {\tilde{z}}^A &=&
{\tilde{ \nabla}}_{\mu} {\tilde {\nabla}}_{\nu} \, z^A
+ \varepsilon \, {\tilde{ \nabla}}_{\mu} {\tilde {\nabla}}_{\nu} \, v^A  + O(\varepsilon^2)\nonumber\\
&=& \nabla_{\mu} \nabla_{\nu} \, z^A  + \varepsilon \, \biggl[\nabla_{\mu} \nabla_{\nu} \, v^A
- {\gamu} \, ^{\lambda}_{\mu \nu} \, \nabla_{\lambda} \, z^A \Biggr] + O(\varepsilon^2) .
\label{twotilde}
\end{eqnarray}
The Riemann tensor induced on the deformed embedding is defined by the same formula
as in the previous section (\ref{GaussCodazzi4}):
\begin{eqnarray}
{\tilde{R}}\, _{\nu \mu \, \lambda  \rho} =
\eta_{AB} \, \biggl[ {\tilde{\nabla}}_{\mu} {\tilde{\nabla}}_{\lambda} \, {\tilde{z}}^A \,
{\tilde{\nabla}}_{\nu} {\tilde{\nabla}}_{\rho} \, {\tilde{z}}^B - {\tilde{ \nabla}}_{\nu} {\tilde{\nabla}}_{\lambda}
\, {\tilde{z}}^A \, {\tilde{\nabla}}_{\mu} {\tilde{\nabla}}_{\rho} \, {\tilde{z}}^B \biggr]
\label{Riodeform}
\end{eqnarray}

Note that in order to calculate the components of the Riemann tensor induced on the deformed manifold
${\tilde{V}}_4$ we use not only the deformed embedding functions ${\tilde{z}}^A$, but
also the ``deformed" covariant derivations ${\tilde{\nabla}}_{\mu}$.

Now, when we insert the expressions like (\ref{twotilde}) into the definition of Riemann tensor components
(\ref{Riodeform}), we shall encounter, besides the zeroth-order initial Riemann tensor
${\displaystyle{{\Rio}}} _{\mu \nu \lambda \rho}$
and the second-order corrections proportional to $\varepsilon^2$, just two types of terms linear in $\varepsilon$:
\begin{eqnarray*}
\quad \varepsilon \, \eta_{AB}\, \nabla_{\mu} \nabla_{\lambda} \, z^A \, \nabla_{\nu} \nabla_{\rho} \, v^B \end{eqnarray*}
and
\begin{eqnarray}
\varepsilon \, \eta_{AB} \,  \nabla_{\mu} \nabla_{\lambda} \, z^A
{\gamu} \, ^{\lambda}_{\nu \rho} \, \nabla_{\lambda} \, z^B .
\label{twoterms}
\end{eqnarray}
The terms of the second type vanish by virtue of the identity (\ref{identity5}); therefore the first-order
correction to the components of Riemann tensor can be written as follows:
\begin{eqnarray}
{\displaystyle{\Riu}}\, _{\nu \mu \lambda \rho} &=& \eta_{AB} \, \biggl[ \nabla_{\mu} \nabla_{\lambda} \, z^A \,
\nabla_{\nu} \nabla_{\rho} \, v^B  +
 \nabla_{\mu} \nabla_{\lambda} \, v^A \,
\nabla_{\nu} \nabla_{\rho} \, z^B\nonumber \\
&-& \nabla_{\nu} \nabla_{\lambda} \, z^A \,
\nabla_{\mu} \nabla_{\rho} \, v^B  - \nabla_{\nu} \nabla_{\lambda} \, v^A \,
\nabla_{\mu} \nabla_{\rho} \, z^B \biggr]
\label{Riemann1}
\end{eqnarray}
To establish the form of linear correction to Einstein's equations we need to know the
components of the first-order correction to the Ricci tensor and the Riemann scalar.
These quantities are readily computed as follows:
\begin{eqnarray}
{\displaystyle{\Riu}}\,_{\mu \rho} = {\displaystyle{\go}}\,^{\nu \lambda} \,
{\displaystyle{\Riu}}\,_{\mu \nu \lambda \rho}
+ {\displaystyle{\gu}}\,^{\nu \lambda} \, {\displaystyle{\Rio}}\,_{\mu \nu \lambda \rho},
\label{Riccifirst}
\end{eqnarray}
Consequently, the first-order correction to the Riemann scalar is:
\begin{eqnarray}
{\displaystyle{\mathop{R}^{1}}} = {\displaystyle{\go}}\,^{\mu \nu} \, {\displaystyle{\Riu}}\, _{\mu \nu}
+  {\displaystyle{\gu}}\,^{\mu \nu} \, {\displaystyle{\Rio}}\, _{\mu \nu} .
\label{Riemscal1}
\end{eqnarray}
Finally, The first-order correction to the Einstein tensor, i.e. the left-hand side of Einstein's equations is:
\begin{eqnarray}
{\displaystyle{\mathop{G}^{1}}}\, _{\mu \nu} &=& {\displaystyle{\mathop{R}^{1}}}\, _{\mu \nu}
- \frac{1}{2} \, {\displaystyle{\gu}}\, _{\mu \nu} \, {\displaystyle{\mathop{R}^{o}}} -
\frac{1}{2} \, {\displaystyle{\go}}\, _{\mu \nu} \, {\displaystyle{\mathop{R}^{1}}}= \\
&=& {\displaystyle{\mathop{R}^{1}}}\, _{\mu \nu}
- \frac{1}{2} \, {\displaystyle{\go}}\, _{\mu \nu} \,
{\displaystyle{\go}}\, ^{\lambda \rho} {\displaystyle{\mathop{R}^{1}}} \, _{\lambda \rho}
- \frac{1}{2} \, {\displaystyle{\go}}\, _{\mu \nu} \, {\displaystyle{\gu}}\, ^{\lambda \rho}
 {\displaystyle{\mathop{R}^{o}}} \, _{\lambda \rho}
- \frac{1}{2} \, {\displaystyle{\gu}}\, _{\mu \nu} \, {\displaystyle{\mathop{R}^{o}}}.\nonumber
\label{Einstein1}
\end{eqnarray}
\indent
In what follows, we shall always suppose that the initial Riemannian manifold is a solution
of Einstein's equations, i.e. an Einstein space which is Ricci-flat and consequently has
zero scalar curvature, too. Therefore the linear correction (of the first order in small
parameter $varepsilon$ ) to the Einstein tensor will reduce to:
\begin{eqnarray}
{\displaystyle{\mathop{R}^{1}}}\, _{\mu \nu}
- \frac{1}{2} \, {\displaystyle{\go}}\, _{\mu \nu} \,
{\displaystyle{\go}}\, ^{\lambda \rho} {\displaystyle{\mathop{R}^{1}}} \, _{\lambda \rho}
\label{Einstein2}
\end{eqnarray}
In the absence of any extra gravitating matter (besides the matter generating the basic solution,
e.g. the central spherical body for Schwarzschild's solution) the equations to solve can be
written in form of a matrix acting on the first-order correction to the Ricci tensor:
\begin{eqnarray}
\Biggl( \delta^{\lambda}_{\mu} \, \delta^{\rho}_{\nu} -
\frac{1}{2} \, {\displaystyle{\go}}\, _{\mu \nu} \,{\displaystyle{\go}}\, ^{\lambda \rho} \Biggr) \;
{\displaystyle{\mathop{R}^{1}}} \, _{\lambda \rho}  = 0
\label{Einstein3}
\end{eqnarray}
But this amounts to the Ricci flatness up to the first order,
because the operator acting on the right on the Ricci tensor in (\ref{Einstein3})
is non-singular; in fact, it is its own inverse:
\begin{eqnarray}
\Biggl( \delta^{\lambda}_{\kappa} \, \delta^{\rho}_{\sigma} -
\frac{1}{2} \, {\displaystyle{\go}}\, _{\kappa \sigma} \,{\displaystyle{\go}}\, ^{\lambda \rho} \Biggr)
\Biggl( \delta^{\kappa}_{\mu} \, \delta^{\sigma}_{\nu} -
\frac{1}{2} \, {\displaystyle{\go}}\, _{\mu \nu} \,{\displaystyle{\go}}\, ^{\kappa \sigma} \Biggr)
= \delta^{\lambda}_{\mu} \, \delta^{\rho}_{\nu}
\label{idempotent}
\end{eqnarray}
>From this we infer that in an Einsteinian background the first-order
correction {\it in vacuo} should satisfy the equation
\begin{eqnarray}
{\displaystyle{\mathop{R}^{1}}} \, _{\lambda \rho}  = 0.
\label{Einstein5}
\end{eqnarray}

This may be written, developing (\ref{Riemscal1}), as:
\begin{eqnarray}
{\Riu}_{\nu\sigma}=
U^{\;\;\;\mu\gamma}_{\nu\sigma\;\;\;A}\nabla_\mu\nabla_\gamma v^A +
V^{\;\;\;\mu}_{\nu\sigma\;\;A}\nabla_\mu v^A =0
\end{eqnarray}
with
\begin{eqnarray}\label{U}
U^{\;\;\;\xi \gamma}_{\nu\rho\;\;\;A} \equiv \eta_{AB}
{\displaystyle{\go}}\,^{\mu\kappa} \left(\delta^\xi _\mu \delta^\gamma
_\rho\nabla_\nu\nabla_\kappa z^B  + \delta^\xi _\nu\delta^\gamma
_\kappa\nabla_\mu\nabla_\rho z^B
 -\delta^\xi _\nu \delta^\gamma _\rho \nabla_\mu\nabla_\kappa z^B
 - \delta^\xi _\mu\delta^\gamma _\kappa\nabla_\nu\nabla_\rho z^B
\right)
\end{eqnarray}
\begin{eqnarray}
V^{\;\;\;\mu}_{\nu\rho\;\;A}&\equiv& \eta_{CD}\eta_{AB}
{\displaystyle{\go}}\,^{\mu\kappa} {\displaystyle{\go}}\,^{\sigma\beta}\left(\delta^\mu_\beta\nabla_\kappa
z^B \nabla_\nu\nabla_\sigma z^C \nabla_\mu\nabla_\rho z^D
 +\delta^\xi_\kappa \nabla_\beta z^B \nabla_\nu\nabla_\sigma z^C
\nabla_\mu\nabla_\rho z^D  \right.\nonumber \\
&-& \left. \delta^\xi_\beta\nabla_\kappa z^B \nabla_\mu\nabla_\sigma
z^C \nabla_\nu\nabla_\rho z^D  - \delta^\xi_\kappa\nabla_\beta z^B
\nabla_\mu\nabla_\sigma z^C \nabla_\nu\nabla_\rho z^D
\right)\label{V}
\end{eqnarray}
In the case when the energy-momentum tensor is present (supposing however that it describes
the influence of matter weak enough in order to keep the basic solution unchanged), one must
use the full Einstein's tensor on the right-hand side. The first correction, linear in $\varepsilon$,
reduces then to only two terms due to the fact that the initial solution is an Einstein space
in vacuo so that ${\displaystyle{\mathop{R}^{0}}} \, _{\lambda \rho}  = 0$ and ${\displaystyle{\mathop{R}^{0}}}$:
\begin{eqnarray}
{\displaystyle{\mathop{R}^{1}}}\, _{\mu \nu}
- \frac{1}{2} \, {\displaystyle{\go}}\, _{\mu \nu} \,
{\displaystyle{\go}}\, ^{\lambda \rho} {\displaystyle{\mathop{R}^{1}}} \, _{\lambda \rho} =
\Biggl[ \delta^{\lambda}_{\mu} \, \delta^{\rho}_{\nu} -
\frac{1}{2} \, {\displaystyle{\go}}\, _{\mu \nu} \,{\displaystyle{\go}}\, ^{\lambda \rho} \Biggr]
\, {\displaystyle{\mathop{R}^{1}}}\, _{\mu \nu}  = - \frac{8 \pi G}{c^4} \, T_{\mu \nu},
\label{Einsteinsource1}
\end{eqnarray}
and this in turn, due to the idempotent property (\ref{idempotent}), can be written equivalently as
\begin{eqnarray}
{\displaystyle{\mathop{R}^{1}}}\, _{\mu \nu}
=  - \frac{8 \pi G}{c^4} \,   \Biggl[ \delta^{\lambda}_{\mu} \, \delta^{\rho}_{\nu} -
\frac{1}{2} \, {\displaystyle{\go}}\, _{\mu \nu} \,{\displaystyle{\go}}\, ^{\lambda \rho} \Biggr]
 \, T_{\lambda \rho}
\label{Einsteinsource2}
\end{eqnarray}
which may prove to be more practical for further calculations especially when the energy-momentum
tensor has a particularly simple form.

\subsection{Second order deformations}

Expanding the deformed embedding functions into the power series of small parameter $\varepsilon$
as earlier seen:

\begin{eqnarray}
{\tilde{z}}^A \, (x^{\mu}) = z^A \, (x^\mu) + \varepsilon \, v^A \,
(x^\mu) + \varepsilon^2 \, w^A \, (x^\mu) + ...,
\label{zetexpansion2B}
\end{eqnarray}

the metric tensor ${\tilde g}_{\mu \nu}$ of the deformed embedding is given by the formula
(\ref{defmetric1}).

The expansion of connection coefficients was already given in eq. (\ref{Chrisexp}).

The second order terms in the expansion of the Riemann tensor are
given by the following formula:
\begin{eqnarray}
{\Rid}_{\mu\nu\sigma\rho}
&=&\eta_{A\,B}\left[\nabla_{\nu}\nabla_{\sigma}z^A\left(\nabla_{\mu}\nabla_{\rho}w^B
- {\gamu}^\kappa_{\mu\rho}\nabla_\kappa v^B -
{\gamd}^\kappa_{\mu\rho}\nabla_{\kappa}z^B \right)\right. + \nonumber \\
&+& \left(\nabla_{\nu}\nabla_{\sigma}v^A -
{\gamu}^\kappa_{\nu\sigma} \nabla_\kappa
z^A\right)\left(\nabla_\mu\nabla_\rho v^B -
{\gamu}^\kappa_{\mu\rho}\nabla_\kappa z^B\right) +\\
&+& \left.\left(\nabla_{\nu}\nabla_{\sigma}w^A -
{\gamu}^\kappa_{\nu\sigma}\nabla_\kappa v^A -
{\gamd}^\kappa_{\nu\sigma}\nabla_{\kappa}z^A
\right)\nabla_{\mu}\nabla_{\rho}z^B - (\mu \leftrightarrow
\nu)\right]\nonumber
\end{eqnarray}
The terms proportional to ${\displaystyle{\gamd}}$ are zero according to
(\ref{identity5}). With a little algebra we find:

\begin{eqnarray}
{\Rid}_{\mu\nu\sigma\rho}
&=&\eta_{A\,B}\left[\nabla_{\nu}\nabla_{\sigma}z^A\nabla_{\mu}\nabla_{\rho}w^B+
\nabla_{\nu}\nabla_{\sigma}w^A\nabla_{\mu}\nabla_{\rho}z^B\right.-\nonumber \\
&-&\left.\nabla_{\mu}\nabla_{\sigma}z^A\nabla_{\nu}\nabla_{\rho}w^B-
\nabla_{\mu}\nabla_{\sigma}w^A\nabla_{\nu}\nabla_{\rho}z^B\right]-\nonumber \\
&-&
{\go}_{\kappa\lambda}{\gamu}^\kappa_{\nu\sigma}{\gamu}^\lambda_{\mu\rho}
+
\eta_{AB}\nabla_{\nu}\nabla_{\sigma}v^A\nabla_{\mu}\nabla_{\rho}v^B+
{\go}_{\kappa\lambda}{\gamu}^\kappa_{\mu\sigma}{\gamu}^\lambda_{\nu\rho}
-
\eta_{AB}\nabla_{\mu}\nabla_{\sigma}v^A\nabla_{\nu}\nabla_{\rho}v^B \nonumber
\end{eqnarray}

The $\displaystyle {\gamu} \, {\gamu}$ terms contain only $z$ and $v$
functions.

To write the second-order correction to Einstein equations we need
the correction to the Ricci tensor:
\begin{eqnarray}
{\Rid}_{\nu\rho} = {\go}\,^{\mu\sigma}{\Rid}_{\mu\nu\sigma\rho} +
{\gu}\,^{\mu\sigma}{\Riu}_{\mu\nu\sigma\rho}+{\gd}\,^{\mu\sigma}{\Rio}_{\mu\nu\sigma\rho}
\end{eqnarray}

with $w$ functions contained only in ${\displaystyle{{\Rid}_{\mu\nu\sigma\rho}}}$ and
in ${\displaystyle{{\gd}\,^{\mu\sigma}}}$.

Looking for vacuum solution we must develop this equation. For the sake of
simplicity we note only that the operators acting on the derivative
of $w^A$ are the same given in (\ref{U}) and (\ref{V}). The know
functions $z^A$ and $v^A$ serve now as the right-hand side of
the equations determining the $w^A$ functions:
\begin{eqnarray*}
{\Rid}_{\nu\rho} = 0 \Longrightarrow U^{\;\;\;\xi\gamma}_{\nu\sigma\;\;\;A}\nabla_\xi\nabla_\gamma w^A +
V^{\;\;\;\xi}_{\nu\sigma\;\;A}\nabla_\xi w^A = B_{\nu\rho}(v^A,z^A)
\end{eqnarray*}

with $B_{\nu\rho}(v^A,z^A)$ combination of derivative of $v^A$ and
$z^A$ functions.

\section{Approximate solutions of Einstein equations}

\subsection{Flat background space-time}

In a Minkowskian space-time $M_4$ parameterized by cartesian coordinates $x^{\mu}= [ct,x,y,z]$
all connection coefficients identically vanish, as well as the components of the Riemann and Ricci tensors.
The flat Minkowskian space can be embedded as a hyperplane in any pseudo-Euclidean space with
more than four dimensions and signature $(1+,(N-1)-)$. Let us choose the simplest case of embedding
in five dimensions:
$$M_4 \rightarrow E^5_{1,4}$$
with the first four components denoting a Minkowskian space-time vector in cartesian coordinates:
\begin{eqnarray}
z^1 = ct, \, z^2 = x, \, z^3 = y, \, z^4 = z, \, \ \ z^5=0,
\label{Minkembed}
\end{eqnarray}
the last cartesian coordinate considered as an extra dimension of $E^5_{1,4}$ orthogonal to the
$M_4$ hyperplane.
All covariant derivatives in (\ref{Riodeform}) can be replaced by partial derivatives, and all
second derivatives of linear embedding functions are identically zero. Therefore in order to investigate
non trivial deformations
of the Minkowskian space embedded as a hyperplane we must go the second order in $\varepsilon$.
This leads to the following equation resulting from the requirement of vanishing Ricci tensor:
\begin{equation*}
{\displaystyle {\Rid}_{\mu \rho}}\ = 0 \Longrightarrow
{\displaystyle{\go}}^{\lambda \nu} \eta_{AB}  \biggl[ \nabla_{\mu} \nabla_{\lambda} \, v^A \,
\nabla_{\nu} \nabla_{\rho} \, v^B
\!-\! \nabla_{\nu} \nabla_{\lambda} \, v^A \,
\nabla_{\mu} \nabla_{\rho} \, v^B  \biggr]=0
\end{equation*}
We shall not consider infinitesimal deformations of the first four coordinates because they coincide with
coordinate transformations in $V_4$; therefore the only non vanishing component of $v^A$ is the remaining
fifth coordinate deformation, expanded in a series of powers of $\epsilon$:

$$z^5 =\epsilon \, v(x^{\mu}) + \epsilon^2 \, w (x^{\mu}) + \epsilon^3 \, h (x^{\mu}) + ...$$
In order to keep the Einstein equations satisfied after deformation up to the second order terms, we must have
\begin{eqnarray}
{\displaystyle {\Rid}_{\mu \rho}\!\! = {\go}}\, ^{\lambda \nu}  \biggl[ \nabla_{\mu} \nabla_{\lambda} \, v \,
\nabla_{\nu} \nabla_{\rho} \, v
- \nabla_{\nu} \nabla_{\lambda} \, v \,
\nabla_{\mu} \nabla_{\rho} \, v  \biggr] = 0
\label{Riemannsquare2}
\end{eqnarray}
Any function of linear combination of cartesian coordinates is an obvious solution of
Eq. (\ref{Riemannsquare2}). Indeed, if we set:
\begin{eqnarray}
v(x^{\mu}) = f\, (k_{\mu} \, x^{\mu} )
\label{planewave}
\end{eqnarray}
inserting the derivatives of $v(k_{\mu} \, x^{\mu})$  into (\ref{Riemannsquare2}) results in
the following simple equation :
\begin{eqnarray}
{\displaystyle{\go}}\, ^{\lambda \nu}  \biggl[ k_{\mu} k_{\lambda} \,
k_{\nu} k_{\rho} \, {v'}^2
- k_{\nu} k_{\lambda}  \,
k_{\mu} k_{\rho} \, {v'}^2  \biggr]
 = k_{\nu}k^{\nu} \, {v'}^2 \, \biggl[ k_{\mu} k_{\rho} - k_{\rho} k_{\mu}
\biggr] = 0,
\label{planekmukmu}
\end{eqnarray}
But in fact, this deformation does not have any physical meaning, because the Riemann tensor, which
is the only observable quantity, identically vanishes:
\begin{eqnarray}
{\displaystyle {\Rid}_{\mu \nu \lambda \rho}} =  \biggl[\,  k_{\mu} \, k_{\lambda} \,
k_{\nu} \, k_{\rho} - k_{\nu} \, k_{\lambda} \, k_{\mu} \, k_{\rho} \, \biggr] = 0
\label{Riemannsquare3}
\end{eqnarray}
The vanishing of the Riemann tensor is not surprising, because the deformation considered
looks like a deformation of a plane into a cylinder, which does not alter its
intrinsic flat geometry.

The fact that there are no wave-like solutions at the first order of deformation of Minkowskian
spacetime suggests that the same situation will prevail when we shall investigate other Einsteinian
manifolds embedded in a pseudo-Euclidan flat space, e.g. the Schwarzschild solution. If the
contrary was true, one could keep the wave-like propagating deformations also in the flat limit,
which would contradict the absence of such solutions among the first-order deformations of the
Minkowskian space-time.

This means that the only hope to produce contributions to the Riemann tensor behaving like
a propagating gravitational field, i.e. the gravitational waves, is to consider the third (and higher)
order deformations of embedded Einsteinian manifolds.
The third order variation for the Riemann tensor in the case
of deformations of all orders orthogonal to the embedded manifold reduces to
the following expression:
\begin{eqnarray}
{\mathop{R}^3}_{\mu\nu\sigma\rho}
&=& \eta_{AB}\left(\nabla_\mu\nabla_\rho v^A \nabla_\nu\nabla_\sigma w^B
+\nabla_\mu\nabla_\rho w^A \nabla_\nu\nabla_\sigma v^B - \right. \nonumber \\
&-& \nabla_\nu\nabla_\rho v^A \nabla_\mu\nabla_\sigma w^B -\left.\nabla_\nu\nabla_\rho w^A
\nabla_\mu\nabla_\sigma v^B\right)\qquad
\label{thirdorder}
\end{eqnarray}

The linear contribution coming from the expressions containing third-order deviation linearly
does vanish because the derivatives of the corresponding $z^5$ coordinate are identically zero.

A wave-like behavior of the Riemann tensor can be produced if we assume that $w$ depends on variables
orthogonal to the worldlines parallel to the vector $k$. For the sake of simplicity,
let us start with the first order deformation in the direction of fifth coordinate, i.e.
orthogonal to the embedded Minkowskian hyperplane $M_4$ as a plane wave propagating along
the $z$-axis:
$$\epsilon \, v( x^{\mu}) = e^{i(\omega t - k z)}.$$

According to our general analysis, by virtue of (\ref{Riemannsquare3}), this deformation does not contribute
to the Riemann tensor, which remains zero even at the second order.
Now let us add up the second order deformation depending on the variables $x$ and $y$ only:

\begin{eqnarray}
z^5 = \epsilon \,  e^{i(\omega t - k z)} + \epsilon^2 \, w (x,y)
\label{deformtwo}
\end{eqnarray}
The only contribution to the third order correction to the Riemann tensor has the form
given by the formula (\ref{thirdorder}) in which the covariant derivatives can be replaced
by partial derivatives given that all Christoffel symbols vanish in cartesian coordinates.
The function $w (x,y)$ must have some non vanishing
second order derivatives; let us make the simplest choice and set $w(x,y) = B \, xy$, with
$B$ = Const having the dimension $cm^{-1}$.

Then the only non vanishing second derivative is $\partial^2_{xy} w = B$. Taking into account
the form of (\ref{thirdorder}), the only non vanishing components are:

\begin{eqnarray}
{\mathop{R}^3}_{\mu x y\rho}
= \left(\partial^2_{\mu y}\, v \, \, \partial^2_{x \rho} \, w
+\partial^2_{\mu y} \, w \, \, \partial^2_{x \rho}\, v - \right.  \left.\partial^2_{x y} \, v \, \, \partial^2_{\mu \rho} \, w
- \partial^2_{x y} \, w \, \, \partial^2_{\mu \rho} \, v \right)
\label{thirdorder2}
\end{eqnarray}
and all other components obtained from this one by permutations of indexes allowed by
the well known symmetries of Riemann's tensor, like e.g. $\displaystyle{{\mathop{R}^3}_{x \mu \rho y}}$, etc.

Now, given that $v$ does not depend on $x$ and on $y$, the only non vanishing term in (\ref{thirdorder2})
is the one containing $\partial_x \partial_y w = B$; so that we have
\begin{eqnarray}
{\mathop{R}^3}_{\mu x y\rho}
=  -  \partial^2_{xy} \, w \, \, \partial^2_{\mu \rho} \, v = - B \, \, \partial^2_{\mu \rho} \, v
\label{thirdorder3}
\end{eqnarray}
There is no contribution to the Ricci tensor coming from ${\displaystyle{\go}^{xy} \, {\mathop{R}^3}_{\mu x y\rho} }$
because the Minkowskian metric tensor is diagonal and ${\displaystyle{\go}^{xy} =0}$; therefore,
to make the Ricci tensor vanish up to the third order means that the following equation
must be satisfied:
\begin{eqnarray}
{\displaystyle{\go}\,^{\mu \rho} \, {\mathop{R}^3}_{\mu x y\rho}  = - B \, \, {\go}\,^{\mu \rho}
\, \, \partial^2_{\mu \rho} \, v = 0}
\label{waveforv}
\end{eqnarray}
This is the wave equation for $v$, imposing the dispersion relation $\omega^2 = c^2 \, k^2$.

The particular form of the "modulating" function $w(x,y)$ can be easily generalized.
As a first step, let us consider an arbitrary quadratic form in variables $x$ and $y$:
let us put
$$w = A \, x^2 + B \, xy + C \, y^2$$
Besides the non vanishing component $\displaystyle{{\mathop{R}^3}_{\mu x y\rho} }$, two
other components of Riemann tensor will appear now:
$$\displaystyle{{\mathop{R}^3}_{\mu x x \rho} = 2 A}, \, \, \\ \, {\rm and} \, \\ \, \, \,
\displaystyle{{\mathop{R}^3}_{\mu y y\rho} = 2 C} $$
which have the same structure as the $(x,y)$ component (\ref{thirdorder3}):
\begin{eqnarray}
{\mathop{R}^3}_{\mu x x \rho}
&=&  -  \partial^2_{x x}\, w \, \, \partial^2_{\mu \rho}\, v = - 2 A \, \, \partial^2_{\mu \rho}\, v,
\, \, \,\nonumber \\
{\mathop{R}^3}_{\mu y y \rho}
&=&  -  \partial^2_{y y}\, w \, \, \partial^2_{\mu \rho}\, v = - 2 C \, \, \partial^2_{\mu \rho}\, v,
\label{thirdorder4}
\end{eqnarray}
The components $(xx), \, (xy) \, \, $ and $(yy)$ of the Ricci tensor vanish if the same condition
(\ref{waveforv}) is satisfied; but now we shall also make sure that all other components
of the Ricci tensor vanish, too, which will be true if the following trace is zero:
\begin{eqnarray}
{\go}\, ^{xx}  {\mathop{R}^3}_{\mu x x \rho} + {\go}\, ^{yy} {\mathop{R}^3}_{\mu yy \rho} &=&  -
\partial^2_{x x}\, w \, \, \partial^2_{\mu \rho}\, v  -  \partial^2_{x x}\, w \, \,
\partial_\mu \partial_\rho v \nonumber \\ &=&
- (2 A + 2 C )\, \, \partial^2_{\mu \rho}\, v
\label{riccitrace}
\end{eqnarray}
leading to the extra condition on the coefficients $A$ and $C$, namely, $A=-C$, thus leaving only
two degrees of freedom for the function $w$. This suggests the {\it quadrupolar} character
of the gravitational wave, which deforms the space simultaneously in two directions perpendicular
to the direction of propagation; notice that if $w$ depended only on one transversal variable,
say $x$, the vanishing of the Ricci tensor would impose $w = 0$ (or a constant, which would not
have any physical meaning at all). It is also worthwhile to note that the fact the planar
wave solutions appear only at the {\it third order} of deformation echoes the well known
result obtained via linearization of the metric tensor, telling that gravitational waves
are emitted when the {\it third} time derivative of the quadrupolar moment is different from
zero.

The same is true for any homogeneous polynomial of two variables $x$ and $y$, provided it
satisfies the two-dimensional Laplace equation $\partial^2_{xx} \, w + \partial^2_{yy} \, w = 0$.
Finally, we can generalize our result by stating that the deformation of Minkowskian space-time
embedded as a hyperplane in an five-dimensional Euclidean ambient space leads to the vanishing
of the Ricci tensor up to the third order in small parameter $\epsilon$ if it has the form
\begin{eqnarray}
z^5 = \epsilon \, e^{i \, (\omega t - k z)} + \epsilon^2 \, w(x,y) + O(\epsilon^3)
\label{wavedeform1}
\end{eqnarray}
Provided that $v$ satisfies $\omega^2 = c^2 k^2$ and $w$ satisfies the two-dimensional Laplace
equation $\nabla^2 w = 0$.

Taking into account that the corresponding Riemann tensor
$\epsilon^3 {\displaystyle{\mathop{R}^3}_{\mu \nu \lambda \rho}} $ is linear both in $v$ and $w$, we can
compose by superposition a transversally polarized plane wave of arbitrary shape and spectrum,
propagating with the phase velocity equal to the speed of light.

The particular form of plane wave solution suggests also the form of a spherical wave.
The first-order deformation far from the source should contain a factor propagating in radial direction,
while the second-order deformation should depend on the angular variables.
We should not expect total vanishing of the second-order correction to the Riemann tensor
like it happened in the case of plane waves. It is important that there will be no propagating
terms at that order of approximation; static terms vanishing at spatial infinity like
$r^{-2}$ or $r^{-3}$ can be neglected and in fact describe the approximation to the static part
of the space-time deformation inevitably produced by the source of spherical gravitational waves.

Let us start with the first-order deformation of Minkowskian space-time embedded
as a hyperplane in some pseudo-Euclidean space; it has one component along one extra dimension perpendicular
to the Minkowskian hyperplane $M_4$. We suppose that is depends on the variables $r$ and $t$ only:
\begin{equation}
v^5 = v^5 \, (t, \, r)
\label{spherdeform1}
\end{equation}
Being perpendicular to the embedded manifold $M_4$ as seen from the host space, this deformation does not
contribute to the first-order correction to the Riemann tensor. In order to evaluate the second-order
correction to the Riemann tensor, ${\displaystyle {\Rid}_{\mu \nu \lambda \rho}}$, we need to insert
the expressions for second covariant derivatives of $v$. In a flat space parameterized by spherical coordinates
the non-vanishing Christoffel symbols are:
\begin{eqnarray*}
\begin{array}{cclcccl}
\Gamma^r_{\theta\theta} &=& -r& \; \; \;&
\Gamma^r_{\phi\phi} &=& -r \sin^2\theta \\
\Gamma^\theta_{r\theta} &=& r^{-1} &\; \; \;&
\Gamma^\phi_{r\phi} &=& r^{-1} \\
\Gamma^\theta_{\phi\phi} &=& -\sin\theta \cos\theta &\; \;&
\Gamma^\phi_{\theta\phi}& =& \cot{\theta}.
\end{array}
\end{eqnarray*}
\noindent
and in the case when $v$ is a function only of $t$ and $r$ the non-vanishing combinations are:

\begin{eqnarray*}
\begin{array}{cclcccl}
\nabla_t \nabla_r  \, v &=& \partial^2_{t r } \, v, &\, \ \ \,& \nabla_t \nabla_t \, v &=& \partial^2_{t t } \, v \\
\nabla_r \nabla_r  \, v &=& \partial^2_{r r } \, v, &\, \ \ \,& \nabla_{\theta} \nabla_{\theta}\, v &=& r \, \partial_r v \\
\nabla_{\varphi} \nabla_{\varphi} \, v &=& r \, \sin^2 {\theta} \partial_r v.
\label{secondderv}
\end{array}
\end{eqnarray*}

We consider the same deformation of the fifth coordinate as before:
$$z^5 =\epsilon \, v(x^{\mu}) + \epsilon^2 \, w (x^{\mu}) + \epsilon^3 \, h (x^{\mu}) + \ldots$$

The first order deformation of the Ricci tensor is still zero.
In order that the the second order be zero we have to satisfy the eq. (\ref{Riemannsquare2}).
Because we are looking for radiative solutions, we shall neglect all terms which decay at spatial
infinity more rapidly than $1/r$, keeping only the radiative part. If we set
\begin{equation*}
v(x^\mu)=\frac{e^{i(\omega t -k r)}}{r}
\end{equation*}

the only  non-vanishing components of radiative character, i.e. behaving at infinity like $1/r$ are:
\begin{eqnarray}
{\displaystyle {\Rid}_{\phi\phi}} = \sin^2\theta {\displaystyle {\Rid}_{\theta\theta}}\sim i k \sin^2\theta \frac{e^{-2i(\omega t -k r)}}{r} \left(k^2 c^2-\omega^2\right)
\end{eqnarray}
and they do vanish provided that $\omega^2=k^2c^2$.
The third order correction to the Riemann tensor is given by the equation (\ref{thirdorder}):
>From this we can easily calculate the third-order correction to the Ricci tensor.
Note that the third order deformation $w$ has to depend on the angles $\theta$ and $\phi$, but must
also have the dimension of a length. This is why we choose
\begin{equation}
w(x^\mu)=r\,\, Q \, (\theta,\phi)
\label{wfunction}
\end{equation}
the only non vanishing components (in radiative approximation) are:
\begin{eqnarray}
{\Rit}_{\phi\phi} = \sin^2 \theta \, {\Rit}_{\theta\theta}\!\sim\!-i k \frac{e^{-i(\omega t -k r)}}{r} \left(\sin^2\theta \;\partial^2_\theta Q + 2 \sin^2\!\theta
\; Q + \partial^2_\phi Q + \cos\theta \sin\theta\; \partial_\theta Q \right).
\label{Ri3}
\end{eqnarray}
Considering the Laplace operator acting on a function (\ref{wfunction}), we see that vanishing of the laplacian
of $w = r \, Q(\theta, \varphi)$ coincides with the condition of vanishing of the two non-trivial
components of the Ricci tensor (\ref{Ri3}):
with respect to the $\theta$ and $\phi$ variables:
\begin{eqnarray}
g^{ij}\nabla_i \nabla_j w(r,\theta,\phi) &=&    \qquad\qquad\qquad\qquad\qquad\qquad \quad\qquad\qquad
 \quad\qquad\qquad         (i,j=\theta,\phi) \nonumber \\
&=&-\frac{1}{r^2}\left(\partial^2_\theta w - \Gamma^r_{\theta\theta} \partial_r w\right)
-\frac{1}{r^2\sin^2\theta}\left(\partial^2_\phi w
- \Gamma^r_{\phi\phi} \partial_r w- \Gamma^\theta_{\phi\phi} \partial_\theta w\right)\nonumber\\
&=&-\frac{1}{r^2\sin^2\theta}\left(\sin^2\theta\; \partial^2_\theta w + 2 r \sin^2\theta\;\partial_r w
+ \partial^2_\phi w + \sin\theta\;\cos\theta\;\partial_\theta w\right)\nonumber\\
&=&-\frac{1}{r\sin^2\theta}\left(\sin^2\theta\; \partial^2_\theta Q + 2 \sin^2\theta\;Q + \partial^2_\phi Q
+ \sin\theta\;\cos\theta\;\partial_\theta Q\right)
\end{eqnarray}
So the third order correction at the Ricci tensor vanish if $w(x^\mu)$ satisfy the
Laplace equation, and the analogy with the plane wave solution we have previously found is complete:

\begin{eqnarray*}
z^5 &=& \epsilon \, \frac{e^{i(\omega t -k r)}}{r} + \epsilon^2 \, w(r,\theta ,\phi) + O(\epsilon^3)\\\; \\
\mbox{with}&& \left\{\begin{array}{ccc} k^2 c^2 &=& \omega^2 \\  \nabla^2 w&=&0 \end{array}\right.
\end{eqnarray*}
The non-radiative terms behaving at infinity like $r^{-2}$ and $r^{-3}$ may give a hint as to the modifications
of Schwarzschild metric that have to be made in order to compensate them thus solving the third-order
Einstein equations exactly, or at least up to that order in the development in powers of $r$. This will
 probably suppose the existence of time-dependent variations of Schwarzschild background that would
serve as the source of our spherical gravitational wave.

\subsection{Isometric embedding of Schwarzschild's solution}
Isometric embeddings of Einstein spaces in pseudo-Euclidean flat spaces
of various dimensions and signatures can be found in J. Rosen's paper in \cite{Rosen1965}.
An embedding of the exterior Schwarzschild solution which is of particular interest to us ,
cited in Rosen's paper, has been found by Kasner \cite{Kasner1921}, who also proved that
the embedding of Schwarzschild's solution in a five-dimensional pseudo-Euclidean space is
impossible. Kasner's embedding uses a pseudo-Euclidean space $E^6$ with signature $(++----)$
and is defined as follows:
\begin{eqnarray}
z^1 &=& MG \Biggl( 1 - \frac{2 MG}{r} \Biggr)^{\frac{1}{2}} \, \cos \biggl( \frac{ct}{MG}\biggr) ,\nonumber \\
z^2 &=& MG \Biggl( 1 - \frac{2 MG}{r} \Biggr)^{\frac{1}{2}} \, \sin \biggl( \frac{ct}{MG}\biggr) , \nonumber \\
z^3 &=&  \int \,  \Biggl[ \frac{1 + \biggl(\frac{MG}{r}\biggr)^4}{1 -
\frac{2MG}{r}} - 1  \Biggr]^{\frac{1}{2}} \, dr , \label{Schwembedding1}\\
z^4 &=& r \sin \, \theta \, \cos \, \phi , \nonumber \\
z^5&=& r \sin \, \theta \, \sin \, \phi , \nonumber \\
z^6 &=& r \, \cos \, \theta \nonumber
\end{eqnarray}

Here $M$ is the mass of the central gravitating body and $G$ denotes Newton's gravitational constant.
(Note the dimensional factor $MG$ in front of the definitions of $z^1$ and $z^2$ in order to give
these coordinates the dimension of length).

The embedded four-dimensional manifold $V_4$ is parameterized by the coordinates $x^{\mu}$, with $\mu = 0,1,2,3$
so that
\begin{eqnarray}
x^0 = ct, \,\, x^1 = r, \,  \, x^2 = \theta, \, \ \  x^3 = \phi
\label{xmu}
\end{eqnarray}

Let us denote the flat metric by
$$\eta_{AB} = diag\, (++----), \, \ \ \, \ \ {\rm with} \, \ \ \, \ \ A,B,... = 1,2,...6.$$

Then it is easy to check that the induced metric on the embedded manifold has indeed the usual
Schwarzschild form:
\begin{eqnarray}
\label{Schwmetric1}
ds^2 = \eta_{AB} \, \partial_{\mu} z^A \, \partial_{\nu} z^B \, d x^{\mu} d x^{\nu}
 = g_{\mu \nu} \, (x^{\lambda}) \, d x^{\mu} d x^{\nu}=\qquad && \\\nonumber \\
= \biggr(1 - \frac{2MG}{r}\biggr) c^2 dt^2 - \frac{dr^2}{\biggr(1 - \frac{2MG}{r}\biggr)}
- r^2(d\theta^2\! -\! \sin^2 \theta d \phi^2)&& \nonumber
\end{eqnarray}

However, this particular embedding is not unique. In $1959$ C. Fronsdal
\cite{Fronsdal1959} proposed a similar embedding into pseudo-Euclidean space with the
signature $(1+, 5-)$, using hyperbolic functions instead the trigonometric ones. Fronsdal's
embedding is defined as follows:
\begin{eqnarray}
z^1 &=& MG \Biggl( 1 - \frac{2 M G}{r} \Biggr)^{\frac{1}{2}} \, \sinh \biggl( \frac{c\,t}{M G}\biggr) ,\nonumber \\
z^2 &=& MG \Biggl( 1 - \frac{2 M G}{r} \Biggr)^{\frac{1}{2}} \, \cosh \biggl( \frac{c\,t}{M G}\biggr) , \nonumber \\
z^3 &=&  \int{ \,\Biggl[ \frac{1 - \biggl(\frac{M G}{r}\biggr)^4}{1 -
\frac{2MG}{r}} - 1  \Biggr]^{\frac{1}{2}} \, }\ud r ,\, \label{Schwembedding2}\\
z^4 &=& r \sin \, \theta \, \cos \, \phi , \nonumber \\
z^5&=& r \sin \, \theta \, \sin \, \phi , \nonumber \\
z^6 &=& r \, \cos \, \theta \nonumber
\end{eqnarray}
Again, it is easy to check that with the pseudo-Euclidean metric $\eta_{AB} = diag(+-----)$
the induced metric
\begin{eqnarray}
g_{\mu \nu} = \eta_{AB} \, \partial_{\mu} z^A \, \partial_{\nu} z^B
\label{Schwmetric2}
\end{eqnarray}
is the same as in (\ref{Schwmetric1}).

We have found a way to encode the two cases in a single formula. Introducing two constants
$\sigma$ and $\chi$ we can write the pseudo-Euclidean six dimensional metric as  $\eta_{AB} = diag(+\, \sigma\, ----)$,
and the embedding functions $z^A(x^\mu)$ as follows:
\begin{eqnarray}\label{Zgeneral}
\begin{array}{ccl}
z^1 &=& MG \Biggl( 1 - \frac{2 MG}{r} \Biggr)^{\frac{1}{2}} \, \frac{\exp \biggl( \frac{ct}{MG} \chi \biggr)-
\chi^2 \exp \biggl( - \frac{ct}{MG} \chi \biggr)}{2} ,\\
z^2 &=& MG \Biggl( 1 - \frac{2 MG}{r} \Biggr)^{\frac{1}{2}} \, \frac{\exp \biggl( \frac{ct}{MG}
\chi \biggr)+\chi^2 \exp \biggl( - \frac{ct}{MG} \chi \biggr)}{2\chi} ,\\
z^3 &=&  \int \,  \Biggl[ \frac{\sigma M^4 G^4 - 2 M G r^3}{(2 M G - r) \, r^3}  \Biggr]^{\frac{1}{2}} \, dr , \\
z^4 &=& r \sin \, \theta \, \cos \, \phi , \\
z^5 &=& r \sin \, \theta \, \sin \, \phi , \\
z^6 &=& r \, \cos \, \theta
\end{array}
\end{eqnarray}

from which it follows that choosing the values $(\sigma=-1, \chi=1)$ one gets Fronsdal's embedding, while by
choosing the values $(\sigma=1, \chi=i)$ one gets Kasner's embedding. However in what follow we will use
Fronsdal's functions.

Let us also write down the non vanishing components of the Christoffel symbols and of the Riemann tensor,
which do not depend on the embedding being inherent to the internal geometry of Schwarzschild's solution.
The Christoffel symbols are the following:
\begin{eqnarray}
\displaystyle{{\gamo}}\, ^{t}_{t r} = \displaystyle{{\gamo}}\, ^{t}_{r t} &=& \frac{MG}{r^2} \,
\biggl( 1 - \frac{2MG}{r} \biggr)^{-1}; \nonumber \\
  \displaystyle{\gamo}\, ^{r}_{r r}  &=& - \frac{MG}{r^2} \,
\biggl( 1 - \frac{2MG}{r} \biggr)^{-1} \\
\displaystyle{\gamo}\, ^{r}_{t t}&=& \frac{MG}{r^2} \,
\biggl( 1 - \frac{2MG}{r} \biggr); \nonumber \\
 \displaystyle{{\gamo}}\, ^{r}_{\theta \theta} &=& - r \,
\biggl( 1 - \frac{2MG}{r} \biggr); \nonumber \\
\displaystyle{\gamo}\, ^{r}_{\varphi \varphi} &=& - r \, \sin^2 \theta \, \biggl( 1 - \frac{2MG}{r} \biggr);\nonumber \\
\displaystyle{{\gamo}}\, ^{\theta}_{\theta r} &=& \displaystyle{{\gamo}}\, ^{\theta}_{r \theta} = \frac{1}{r}; \nonumber \\
\displaystyle{{\gamo}}\, ^{\theta}_{\varphi \varphi } &=& - \sin \theta \, \cos \theta; \nonumber \\
\displaystyle{{\gamo}}\, ^{\varphi}_{\varphi r } = \displaystyle{{\gamo}}\, ^{\varphi}_{r \varphi}
&=& \displaystyle{\frac{1}{r}}; \nonumber \\
\displaystyle{\gamo}\, ^{\varphi}_{\varphi \theta } &=& \displaystyle{\gamo}\, ^{\varphi}_{\theta \varphi} =
\frac{\cos \theta}{\sin \theta}.
\label{ChristoffelsS}
\end{eqnarray}

while the non vanishing components of Riemann's tensor of Schwarzschild's metric are given by the following expressions:

\begin{eqnarray}
{\displaystyle{{\Rio}}}_{\, t \,  r \, t \, r}  &=&  \frac{2MG}{r^3}\nonumber \\
{\displaystyle{{\Rio}}}_{\, t \,  \theta \, t \, \theta}  &=& - \frac{MG}{r} \biggl(1 - \frac{2MG}{r} \biggr) \nonumber \\
{\displaystyle{{\Rio}}}_{\, t \,  \varphi \, t \, \varphi}  &=& - \frac{MG}{r} \biggl(1 - \frac{2MG}{r} \biggr) {\sin}^2
\, \theta\nonumber \\
{\displaystyle{{\Rio}}}_{\, r \,  \theta \, r \, \theta}  &=& \frac{MG}{r} \biggl(1 - \frac{2MG}{r} \biggr)^{-1} \\
{\displaystyle{{\Rio}}}_{\, r \,  \varphi \, r \, \varphi}  &=& \frac{MG}{r} \biggl(1 - \frac{2MG}{r} \biggr)^{-1}
\, {\sin}^2 \, \theta \nonumber \\
{\displaystyle{{\Rio}}}_{\, \theta \,  \varphi \, \theta \, \varphi}
&=& - 2MGr \, {\sin}^2 \, \theta \nonumber
\end{eqnarray}

\subsection{Deformations of the embedded Schwarzschild manifold}

As explained above, the transversality condition (\ref{transversality1}) which represents
four independent equations, shall leave only two arbitrary functions describing non-trivial
deformations of the exterior Schwarzschild solution in six-dimensional pseudo-Euclidean space.
Let us find as simple choice as possible in order to make easier the subsequent calculus of
Einstein equations.

Let us examine the transversality conditions (\ref{transversality1}) one by one in the concrete
case of the embedding given by formulae (\ref{Schwembedding1}). Only the first two embedding
functions depend on time $t$; therefore the $t$-component of transversality condition becomes
\begin{eqnarray}
v^1 \, \partial_0 \, z^1  + v^2 \, \partial_0 \, z^2  = 0
 =- \biggl( 1 - \frac{2MG}{r} \biggr)^{\frac{1}{2}} \,
\biggl[ v^1 \sin \biggl( \frac{ct}{MG} \biggr)  - v^2 \cos \biggl( \frac{ct}{MG} \biggr)  \biggr]
\label{gauge12}
\end{eqnarray}
It follows that the two functions $v^1$ and $v^2$ are proportional to a common function $v(x^{\mu})$,
and the ansatz that solves (\ref{gauge12}) is:
\begin{eqnarray}
v^1 &=& v(x^{\mu}) \, z^1 = MG \; v(x^\mu) \; \sinh{\left(\frac{ct}{MG}\right)},\nonumber\\
v^2 &=& v(x^{\mu}) \, z^2 = MG \; v(x^\mu) \; \cosh{\left(\frac{ct}{MG}\right)}.\label{vcommon}
\end{eqnarray}

The common factor $MG$ in front of the definitions above is put there to give the proper dimension
(length) to $v^1$ and $v^2$. The unknown function $u$ is then dimensionless.
By analogy one can observe that the only embedding functions which depend on angular variables $\theta$ and $\phi$
are the last three ones,  $z^4, \, z^5$ and $z^6$. Therefore the simplest way to satisfy simultaneously the last two
equations corresponding to the components $\theta$ and $\phi$ of (\ref{transversality1}) is to set
\begin{eqnarray}
v^4 \sim z^4, \, \ \, v^5 \sim z^5, \, \ \, v^6 \sim z^6 ,
\label{v456}
\end{eqnarray}
or more explicitly,
\begin{eqnarray}
v^4 &=& \frac{MG}{r}\,w(x^{\mu}) \, z^4 = MG\; w(x^{\mu}) \, \sin \, \theta \, \cos \, \phi , \,\nonumber \\
v^5 &=&  \frac{MG}{r}\,w(x^{\mu}\,)z^5 = MG \;w(x^{\mu}) \, \sin \, \theta \, \sin \, \phi , \, \nonumber \\
v^6 &=&  \frac{MG}{r}\,w(x^{\mu}) \, z^6 = MG \;w(x^{\mu})  \, \cos \, \theta. \nonumber
\label{v456exp}
\end{eqnarray}
Again, we have conserved the functions $z^A$ in these formulae in order to maintain the
spatial dimension in the definition of embedding functions and their deformations, while
the unknown functions $v$ and $w$ will be kept dimensionless.

Now we have two independent functions of four-dimensional coordinates, $u (x^{\mu})$ and $w(x^{\mu})$.
The only remaining component of $v^A$ which we shall denote by $v^3 = (MG) \, h(x^{\mu})$,
is entirely determined by the radial component of the transversality condition (\ref{transversality1}),
$\eta_{AB} \, v^A \partial_r \, z^B = 0$. It reads:
\begin{eqnarray*}
 M^2G^2 v(x^\mu) +h(x^\mu) r^2\;\Biggl[ \frac{ (1 - \frac{MG}{r} )^4 }{ ( 1 - \frac{2MG}{r} )}
- 1 \Biggr]^{\frac{1}{2}}+ \, r^2\;w(x^\mu) = 0.
\end{eqnarray*}
After simplifying by the common factor $(MG)$ we can express the function $h(x^{\mu})$ as
the following linear combination of $v$ and $w$:
\begin{eqnarray*}
h (r) = - \frac{\left( M^2G^2 v(x^\mu) + r^2\;w(x^\mu)\right)}{r^2\left[\frac{MG}{r^3}
\left(\frac{2r^3-M^3G^3}{r-2MG}\right)\right] ^{\frac{1}{2}}}
\end{eqnarray*}
So finally we can write
\begin{eqnarray}
v^1 &=&  MG \; v(x^\mu) \; \sinh{\left(\frac{ct}{MG}\right)},\nonumber\\
v^2 &=&   MG \; v(x^\mu) \; \cosh{\left(\frac{ct}{MG}\right)}\nonumber \\
v^3 &=& - M G \frac{\left( M^2G^2 v(x^\mu) + r^2\;w(x^\mu)\right)}{r^2\left[\frac{MG}{r^3}
\left(\frac{2r^3-M^3G^3}{r-2MG}\right)\right] ^{\frac{1}{2}}}\\
v^4 &=& MG\; w(x^{\mu}) \, \sin \, \theta \, \cos \, \phi , \,\nonumber \\
v^5 &=&  MG \;w(x^{\mu}) \, \sin \, \theta \, \sin \, \phi , \, \nonumber \\
v^6 &=& MG \;w(x^{\mu})  \, \cos \, \theta. \nonumber
\end{eqnarray}

\subsection{First order approximation of Einstein equations}

\indent
The system (\ref{Einsteinsource1}) represents an approximate version of the full system
of Einstein equations with a source corresponding to the presence of extra matter
in the vicinity of the central spherically symmetric mass. The extra matter gives
rise to the energy-momentum tensor which can be that of of a test particle turning around the central body,
or to an axially-symmetric distribution of matter, e.g. a homogeneous disc turning
around the central body in the equatorial plane. It can be also a spherically symmetric electromagnetic
field defined everywhere in the space surrounding the central body. The last solution to the
problem is well know, it is the Reissner-Nordstr{\o}m metric generated by a spherically symmetric
mass endowed with an electric charge $Q$.

In all these cases we should suppose that the absolute value of the energy-momentum tensor
is very small when compared to the central mass $M$; in other words, the total mass $m$ of the
rotating disc or that of the point-particle representing a satellite should be small enough in order
to treat the ration $m/M$ as an infinitesimal parameter of the deformation of the background space-time
provoked by its presence. All non-vanishing components of Riemann curvature tensor (and those
of the Ricci tensor, too) in the case of Schwarzschild metric are proportional to $M$; {\it in vacuo}
Schwarzshild metric is Ricci-flat and its Einstein tensor vanishes; but if the right-hand side
is not zero anymore but proportional to $m$, Einstein's equations can be divided on both
sides by $M$, and the right-hand side will become proportional to the dimensionless small parameter
$\varepsilon = (m/M)$

If a solution of the linearized problem can be found, it will automatically generate second and higher
order corrections on both sides. One should keep in mind the fact that the energy-momentum tensor
chosen as the right-hand side of Einstein's equations must be also a solution of the systam
of equations ruling the motion of matter in a given background, e.g. the geodesic equation when
one wants to describe the motion of a satellite with a very small mass in a given space-time
geometry. But if this geometry changes (in our case via a perturbation proportional to the
small parameter $\varepsilon$), so should also the geodesic equations, because the connection coefficients
have been modified by addition of small perturbation proportional to $\varepsilon$. The
geodesic equation thus deformed will lead to slightly modified solutions, the unperturbed one
plus the perturbation linear in $\varepsilon$. The source changed in this way, the right-hand side
of Einstein's equations would contain now the terms proportional to $m/M$ and the new ones
proportional to $\varepsilon \, (m/M) = (m/M)^2$, and the second-order approximation of
Einstein's equations should be solved, too, and so forth.

Before trying to find a solution with a non vanishing right-hand side we should determine
the solutions of the homogeneous system (\ref{Einstein5}). There is one solution to the
problem which is quite obvious: the functions $v^A \, (x^{\mu})$ corresponding to an
infinitesimal variation of the mass parameter, $M \rightarrow M + \delta M$. The new embedding
defines just a new Schwarzschild solution with a slightly different mass. We have
 \begin{eqnarray*}
 z^A\,(x^\mu, M+\delta M) = z^A(x,M) + \frac{\delta M}{M} \frac{M \partial z^A}{\partial M} +
\ldots \equiv&& \nonumber \\
z^A(x,M) + \epsilon v^A(x) + \ldots \qquad &&\end{eqnarray*}
where we have defined
$$\epsilon \equiv \frac{\delta M}{M}\ll 1$$ and
\begin{eqnarray}
v^A \, (x^{\mu}) \equiv M \frac{\partial z^A}{\partial M}.
\label{massvariation1}
\end{eqnarray}

By differentiating the embedding functions with respect to the mass
parameter $M$ we find (\ref{Zgeneral})
\begin{eqnarray}\label{Vmassa}
v^1 \, (x^\mu)&=& M G \frac{1-\frac{3MG}{r}}{\sqrt{1- 2 \frac{M G}{r}}} \sinh{\frac{ct}{MG}} -
\sqrt{1- 2 \frac{M G}{r}}\;\; c t\; \cosh{\frac{c t}{M G}}\nonumber \\
&&  \nonumber \\
v^2 \, (x^\mu)&=& M G \frac{1-\frac{3MG}{r}}{\sqrt{1- 2 \frac{M G}{r}}} \cosh{\frac{ct}{MG}} -
\sqrt{1- 2 \frac{M G}{r}}\;\; c t\; \sinh{\frac{c t}{M G}}\nonumber \\
&& \nonumber \\
v^3\, (x^\mu)&=& \int \, \frac{M G \,r^4 + 3 M^5 G^5 \,- 2 M^4 G^4 \,r}{r^3 (-2 M G + r)^2\sqrt{\frac{2 M G r^3
- M^4 G^4}{r^3(-2 M G + r)}}}\,dr  \label{V6massa}\\
&& \nonumber \\
v^4\, (x^\mu)&=& 0\nonumber\\ && \nonumber \\
v^5\, (x^\mu)&=& 0\nonumber\\&& \nonumber \\
v^6\, (x^\mu)&=& 0\nonumber
\end{eqnarray}
In order to find other solutions we have to consider the ten equations (\ref{Einstein5}).
We search for $v(x^\mu)$ and $w(x^\mu)$ on the form
$$v(t,r,\theta,\phi)= f_v(\theta) g_v(r) h_v(\phi) k_v(t)$$
$$w(t,r,\theta,\phi)= f_w(\theta) g_w(r) h_w(\phi) k_w(t) .$$

Consider first the $\theta \phi$ component of $\displaystyle \Riu$

\begin{eqnarray}\label{Ricci34}
\begin{array}{ccl}
\displaystyle{\Riu} \,_{\theta\,\phi} &=& (- M^3 G^3 r^2 - 2 r^5)^{-1} \times \nonumber \\
&\times&\bigg( 2 \, r \cot\theta
(- \,r^4 + M^3 G^3 r - 3 M^4 G^4) \partial_{\phi} v \\
 &\,&+ M G \cot \theta \left[-r^4  - M^3 G^3 ( 3 M G -2 r)\right] \partial_{\varphi}  w \\
 &\,& -2\, r  (-r^4 + M^3 G^3 r - 3 M^4 G^4) \partial^2_{ \varphi \, \theta} v \\
 &\,&- M G  \left[-r^4  - M^3 G^3 ( 3 M G -2 r)\right] \partial^2_{ \varphi \, \theta} w \bigg)
\end{array}
\end{eqnarray}

this expression will vanish automatically if we set
\begin{eqnarray}\label{thetaf}
f_v(\theta) = f_w(\theta)= \sin \theta
\end{eqnarray}
With this choice we get the following expressions for the $t\theta$ and $t\,\phi$ components
\begin{eqnarray}\label{Ricci13}
\Riu \,_{t\, \theta} &=& -\frac{M G \cos \theta}{c r^2(2 r^3 - M^3 G^3)}\times \qquad\quad\quad\quad
\quad\quad\quad\quad \quad\quad \nonumber \\
&\times&\biggl( 3 r^2 M^2 G^2 \left(1-\frac{2 M G}{r}\right) \,g_v(r)\,h_v(\phi) \,k_v '(t)\nonumber \\
&+& (r^4 - 3 M^4 G^4 + M^3 G^3 r) g_w(r) \, h_w(\phi) \; k_w ' (t) \biggr)
\end{eqnarray}
\begin{eqnarray}\label{Ricci14}
\Riu \,_{t\,\phi} &=& - \frac{M G \sin \theta}{c r^2(2 r^3 - M^3 G^3)}\times \qquad\quad\quad\quad
\quad\quad\quad\quad \quad\;\;\quad \nonumber \\
&\times& \biggl( 3 r^2 M^2 G^2 \left(1-\frac{2 M G}{r}\right) \,
\, g_v(r)\,h_v '(\phi) \,k_v '(t)\nonumber \\
&+& (r^4 - 3 M^4 G^4 + M^3 G^3 r)\,g_w(r)\, h_w '(\phi) \,k_w ' (t) \biggr)
\end{eqnarray}

They are both equal to zero if we set

\begin{eqnarray}\label{phi t r}
\begin{array}{ccl}
k_v (t) &=& k_w (t) \\
&\,&\\
h_v (\phi) &=& h_w(\phi)\\
&\,&\\
g_v(r) &=& - \frac{r^4 + 3 M^4 G^4 \sigma - M^3 G^3 r \sigma}{3 M^2 G^2 (2 M G - r) r \sigma} g_w (r)
\end{array}
\end{eqnarray}

The components $r\, \theta$ and $r\,\phi$ can be now written as

\begin{eqnarray}\label{Ricci23}
\begin{array}{ccl}
{\displaystyle{\Riu}} _{r\,\theta} &=& - \frac{\cos \theta \, k_w(t)\, h_w (\phi)}{3 M^2 G^2 r^2 (r- 2 M G)^2}\,
\Bigg[\biggl( M^4 G^4 - 2 M G\,r + r^4\biggr)  \\
&\quad& \biggl( (3 M G - r) g_w(r) +(2 M G-r)\,r g_w ' (r) \biggr)\Bigg]
\end{array}
\end{eqnarray}
\begin{eqnarray}\label{Ricci24}
\begin{array}{ccl}
{\displaystyle{\Riu}} _{r\,\phi} &=& - \frac{\sin \theta \, k_w(t)\, h_w '(\phi)}{3 M^2 G^2 r^2 (r- 2 M G)^2}\,
\Bigg[\biggl( M^4 G^4 - 2 M G\,r + r^4\biggr)  \\
&\quad& \biggl( (3 M G - r) g_w(r) +(2 M G-r)\,r g_w ' (r) \biggr)\Bigg]
\end{array}
\end{eqnarray}
which are both equal to zero if we set

\begin{eqnarray}
g_w(r)= \frac{\sqrt{1- 2\frac{M G}{r}}}{r}
\end{eqnarray}

Finally, the diagonal components are all zero if for the remaining functions we set:

\begin{eqnarray*}
\begin{array}{ccl}
h_w(\phi) &=&  a \cos \phi + b \sin \phi  \\
k_w(t) &=& A \frac{ e^ { \frac{ct}{MG}\, \chi} - \chi^2 e^{- \frac{ct}{MG}\, \chi }}{2}+
B \,\frac{e^{\frac{ct}{MG} \chi }+ \chi^2 e^{- \frac{ct}{MG}\, \chi }}{2 \chi} \\
\end{array}
\end{eqnarray*}
So finally we can write the deformation functions as:
\begin{eqnarray}
v(x^\mu)&=& M G\; \frac{r^4 - 3 M^4 G^4 + r\,M^3 G^3}{3 r^2\,M^2 G^2 \left(2 M G - r\right)}
\sqrt{1- \frac{2 M G}{r}} \sin{\theta}\times \nonumber \\
&\times&  \left(a \sin{\phi} + b \cos{\phi}\right)\left(A \sinh{\frac{c\,t}{M\,G}} + B \cosh{\frac{c\,t}{M\,G}}\right)
\nonumber \\
w(x^\mu) &=& M G \frac{\sqrt{1- \frac{2 M G}{r}}}{r} \sin{\theta} \left(a \sin{\phi} + b \cos{\phi}\right)
\times \nonumber \\
&\times& \left(A \sinh{\frac{c\,t}{M\,G}} + B \cosh{\frac{c\,t}{M\,G}}\right) \label{secondadef}
\end{eqnarray}

However, there is an essential difference between these two solutions. While the first one describes
Riemann tensor whose all non vanishing components tend to zero at space infinity as $r^{-3}$, the second one,
although strictly Ricci-flat, has some components of Riemann tensor which do not vanish at space infinity;
in the case when the hyperbolic functions are chosen we have:

\begin{eqnarray*}
{\Riu}_{t\,\theta\,t\,\theta} \quad \mathop{\longrightarrow}_{r\rightarrow \infty} &&
- 2 M G \sin{\theta} \cos{\phi} \sinh{\frac{ct}{MG}} \\
{\Riu}_{t\,\phi\,t\,\phi} \quad\mathop{\longrightarrow}_{r\rightarrow \infty} &&
- 2 M G \sin^3{\theta} \cos{\phi} \sinh{\frac{ct}{MG}} \\
{\Riu}_{\phi\,\theta\,\phi\,\theta} \quad\mathop{\longrightarrow}_{r\rightarrow \infty} &&
6 M^3 G^3 \sin^3{\theta} \cos{\phi} \sinh{\frac{ct}{MG}} \\
{\Riu}_{t\,\phi\,\theta\,\phi} \quad\mathop{\longrightarrow}_{r\rightarrow \infty} &&
- 3 M^2 G^2 \cos{\theta} \sin^2{\theta} \cos{\phi} \cosh{\frac{ct}{MG}} \\
{\Riu}_{t\,\theta\,\phi\,\theta}  \quad\mathop{\longrightarrow}_{r\rightarrow \infty} &&
3 M^2 G^2\cos{\theta} \sin{\theta} \sin{\phi} \cosh{\frac{ct}{MG}}
\end{eqnarray*}

Such a situation occurs in electrodynamics, too, where the constant electric or magnetic field
tensor obviously satisfies Maxwell's equations, but can not be considered as a regular solution
at infinity. Nevertheless such field can be used as an independent solution together with a regular
solution, e.g. the Coulomb-like field, in a bounded portion of space.

In our case it is clear that the only solution acceptable at space infinity is the first one,
generated by infinitesimal mass variation. This is similar to what occurs in classical
electromagnetism: at very great distances all static fields generated by finite distribution
of charges look at the first approximation as a Coulomb potential of the total charge; then
comes the potential of a dipole momentum, etc.

Having this second independent solution by no means could span the space of all solutions of our linear system.
In the case of partial differential equations the space of solutions is infinite-dimensional,
and in order to single out a solution one has to specify the initial or boundary conditions
on an entire three-dimensional submanifold.

Nevertheless the existence and the character of the two independent solutions of Einsteinian deformations
of the Schwarzschild embedding, we conclude
that any small perturbation of Schwarzschild solution, confined in a finite space surrounding the central
body, is seen at infinity just as a Schwarzschild metric corresponding to the sum of the two masses,
$M + m$.

In a four dimensional space-time the above two solutions can coexist only separated by a three
dimensional manifold dividing the four dimensional manifold in two disconnected parts.
Such a discontinuity cannot correspond to a point-like object moving in the vicinity of the
central mass along the worldline given by $\stackrel{0\;\;}{x^\mu}(s)$,
because the Dirac delta function describing such an object with the following
energy-momentum tensor:

\begin{eqnarray}
T^{\mu\nu} = m\, \delta^4\!\!\left(x^\sigma - \stackrel{0\;\;}{x^\sigma}\!(s)\right)
u^\mu(x^\sigma)\,u^\nu(x^\sigma)
\label{Energiaimpulso}
\end{eqnarray}

where the four-dimensional delta function means literally the tensor product:

\begin{eqnarray*}
\delta^4\!\!\left(x^\sigma - \stackrel{0\;\;}{x^\sigma}\!(s)\right)\equiv
\delta(t-\stackrel{0}{t}(s))\,\delta(r-\stackrel{0}{r}(s))\, \delta(\theta-\stackrel{0}{\theta}(s))
\;\delta(\phi-\stackrel{0}{\phi}(s)),
\end{eqnarray*}
with $u^\mu(x^\sigma)$ denoting the components of the four-velocity of the mass $m$.

In this case the discontinuity is concentrated on a one-dimensional line which cannot divide the
four dimensional space-time in two disconnected parts.

However, the two solutions can be separated by e.g. a spherical surface which in the space-time
is multiplied by the time axis, thus separating the space-time in two disconnected parts by
means of Heaviside's function:

\begin{equation}
H(r-R) = \left\{
\begin{array}{ccc}
1 &\quad& \mbox{ for }r>R\\
0 &\quad& \mbox{ for }r<R
\end{array} \right.
\end{equation}

It is clear that far from the source only the well behaving solution survives,
corresponding just to the variation of central mass without changing the symmetry.
Other deformations can be found only if a definite symmetry breaking is chosen.

\section{Second order deformations and axial symmetry}

\subsection{General considerations}

The most important difference between the first and all the subsequent orders of
deformation of embedded manifolds is that one can no more neglect the possibility of
entering extra euclidean dimensions. The initial embedding in $E^N_{(p,q)}$ can
always be considered as a hyperplane in some higher-dimensional Euclidean space,
$E^{(N+n)}_{(p',q')}$, with $p'+q'-p-q = n$. As it was shown in Section 3.1.,
the first order deformations going beyond this hyperplane do not contribute to
the modifications of Einstein's equations. But starting from the second order in
infinitesimal parameter $\varepsilon$ the deformations along the initial hyperplane
and those from the extra dimensions will contribute simultaneously to the second-order
modifications to the Einstein equations, and the number of unknown functions to be found
will become higher than before.

The Kerr-Newman metric in Schwarzschild-like coordinates (see \cite{BoyerLindquist1967})
can be written as follows:

$$ ds^2 = \biggl( 1 - \frac{2MGr}{\Sigma} \biggr) \, c^2 dt^2
+\frac{4 a MG r \sin \theta}{\Sigma} \, c dt d\varphi - \frac{\Sigma}{\Delta} dr^2$$
\begin{eqnarray}
- \Sigma \, d \theta^2 -
\biggl( \Delta + \frac{2 MG r (r^2 + a^2)}{\Sigma} \biggr) \, {\sin}^2 \theta d \varphi^2
\label{Kerrmetric}
\end{eqnarray}
where
\begin{eqnarray}
\Sigma = r^2 + a^2 \, {\cos}^2 \theta, \, \ \ \, \, \ \ \Delta = r^2 - 2 \, MGr + a^2.
\label{Sigmadelta}
\end{eqnarray}

It tends to the square of the Schwarzschild line element when $a \rightarrow 0$.

It is natural to ask whether this metric can be also embedded $E^6_{(1,5)}$, as was the
case of the exterior Schwarzschild solution. One cannot exclude {\it a priori} such
a possibility; the Reissner-Nordstr{\o}m metric

\begin{eqnarray*}
ds^2 = \biggr(1 - \frac{2MG}{r} + \frac{Q^2 G}{4 \pi \varepsilon_0  \, r^2} \biggr) \, c^2 d t^2 -
\qquad\qquad\qquad&&    \nonumber \\
-\biggr(1 - \frac{2MG}{r} + \frac{Q^2 G}{4 \pi \varepsilon_0  \, r^2} \biggr)^{-1} \, dr^2
 - r^2 d \theta^2 - r^2 \sin^2 \theta d \phi^2 &&
\label{ReissnerNord1}
\end{eqnarray*}

can be embedded in $E^6_{1,5}$
in the same manner as the Schwarzschild solution, but this is due to its spherical
symmetry.

\subsection{The Kerr metric as a deformation of Schwarzschild background}

The axially symmetric Kerr-Newman solution can not be embedded in six (pseudo)Euclidean dimensions;
it requires more dimensions than that, for the following reason. Whatever the dimension of
ambient flat space, the embedding functions for this metric must depend on two extra parameters, $?$ and $a$.
For each value of these two  parameters we have a non-trivial four-dimensional Riemannian manifold embedded
in a flat space; but if we vary $M$ and $a$ independently, we get a two-parameter congruence of embedded
four-dimensional manifolds, which can be looked upon  as a six-dimensional manifold, which is by no
means flat. This can be checked by calculus of the Riemann tensor of the corresponding six-dimensional
metric. Therefore  a higher than six-dimensional flat space is required for the embedding of this non trivial
six-dimensional manifold.

In the early eighties R.R. Kuzeev has defined an isometric embedding of Kerr's metric
(\ref{Kerrmetric}) in a nine dimensional pseudo-Euclidean space (\cite{Kuzeev1981}).

Before finding the embedding, it is much easier to define {\it four} Pfaffian one-forms
whose squares summed up with appropriate signs will give the desired line element in
diagonal form:
\begin{equation}
ds^2 = {\omega}_0^2  -{\omega}_1^2 - {\omega}_2^2 -{\omega}_3^2,
\label{Kerrpfaff}
\end{equation}
with 

$$\omega_0  = \Biggl( 1 - \frac{2MGr}{\Sigma} \Biggr)^{\frac{1}{2}} c dt + 
\frac{2MGra \sin^2 \theta}{\Sigma} \, \Biggl( 1 - \frac{2MGr}{\Sigma} \Biggr)^{-\frac{1}{2}} \, d \varphi , $$
\begin{equation}
\omega_1  = \sqrt{\frac{\Sigma}{\Delta}} \,  dr, \; \; \omega_2 = \sqrt{\Sigma} \, d \theta, \; \;
 \omega_3 = \sqrt{\Delta}  \Biggl( 1 - \frac{2MGr}{\Sigma} \Biggr)^{-\frac{1}{2}} \sin \theta \, dr.
\label{Kerrforms}
\end{equation}
\noindent
with (\ref{Sigmadelta}) defined as above.

It is easy to see that the squares of these differential expressions yield the Kerr-Newman metric
when inserted in the formula (\ref{Kerrpfaff}), and give the Schwarzschild line element
in the limit $a \rightarrow 0$. 

The embedding problem can be formulated now in terms of the integrability of these forms. One should
find $4N$ functions $Q^{A \, \mu} (x^{\lambda})$ satisfying the following identities:
\begin{equation}
\frac{\partial z^A}{\partial x^{\mu}} \, dx^{\mu} = Q^{A\,\mu} (x^{\lambda}) \omega_{\mu}
\label{integable}
\end{equation}

The result is as follows:

\begin{eqnarray}
z^1 &=& (MG) \, \Biggl[ 1 - \frac{2MG}{\Sigma} \, \biggl( 1 - \frac{a}{MG} \, \sin^2 \theta\biggr)
\Biggr]^{\frac{1}{2}} \, {\rm sh} \biggl(\frac{ct}{MG} \biggr),\nonumber\\
z^2 &=& (MG) \, \Biggl[ 1 - \frac{2MG}{\Sigma} \, \biggl( 1 - \frac{a}{MG} \, \sin^2 \theta\biggr)
\Biggr]^{\frac{1}{2}} \, {\rm ch} \biggl(\frac{ct}{MG} \biggr),\nonumber\\
z^3 &=& \Phi_1 \, (r, \theta),\nonumber\\
z^4 &=& \Biggl[ r^2 + a^2 - \frac{2M^2G^2 a r}{\Sigma} \, \biggl(1 - \frac{a}{MG} \,
\sin^2 \theta \biggr) \Biggr]^{\frac{1}{2}} \, \sin \theta \cos \varphi,\nonumber\\
z^5 &=& \Biggl[ r^2 + a^2 - \frac{2M^2G^2 a r}{\Sigma} \, \biggl(1 - \frac{a}{MG} \,
\sin^2 \theta \biggr) \Biggr]^{\frac{1}{2}} \, \sin \theta \sin \varphi,\nonumber \\
z^6 &= &\Biggl[ r^2 + a^2 - \frac{2M^2G^2 a r}{\Sigma} \, \biggl(1 - \frac{a}{MG} \,
\sin^2 \theta \biggr) \Biggr]^{\frac{1}{2}} \, \cos \theta , \nonumber\\
z^7 &=& a \, \Phi_2 \, (r, \theta),
\label{Kuzeevembedding}\\
z^8 &=& ( MG) \, \Biggl( \frac{2ar}{\Sigma} \Biggr)^{\frac{1}{2}} \, \sin \theta \,
\sin \biggl( \varphi - \frac{ct}{MG} \biggr) ,\nonumber\\
z^9 &=& ( MG) \, \Biggl( \frac{2ar}{\Sigma} \Biggr)^{\frac{1}{2}} \, \sin \theta \,
\cos \biggl( \varphi - \frac{ct}{MG} \biggr)\nonumber
\label{Kerrembed}
\end{eqnarray}

with functions  $\Phi_1 \, (r, \theta)$ and $\Phi_2 \, (r, \theta)$ defined by quite
complicated integrals.

When the angular momentum parameter $a$ tends to zero, we recover the Schwarzschild metric
embedded in a six-dimensional pseudo-Euclidean space, the last three embedding functions
disappearing with $a =0$. Yowever, the limit is not analytic because of the presence of
the square root of $a$ in the formulas (\ref{Kerrembed}), and radically different topology
of the two solutions. Nevertheless we believe that this embedding may serve as a starting point for
deformations describing non axially symmetric departures from the Kerr-Newman metric.
This will be the subject of subsequent papers.

\section{Discussion and conclusions}

In this article we have set forth a new formalism that makes it easier to consider small perturbations
of a given Einsteinian background without unphysical degrees of freedom that mar traditional
computations based on the deformations of the metric. Here the embedding provides us with clear
geometric criterion selecting physical degrees of freedom and eliminating the unphysical ones.

We have succeeded the construction of wave solutions in flat space, or in an asymptotically flat
Schwarzschild manifold at spatial infinity. These solutions can display any imposed form due to
the possibility of linear superposition of Legendre polynomials. Once a radiative solution is chosen,
we can extrapolate it towards the smaller values of $r$ where the non-radiative terms prevail.
These can be seen as the corrections to Schwarzschild metric close to the central body that are
responsible for the emission of gravitational waves detected at spatial infinity.

The treatment of this problem is the subject of the work in progress.

\vskip 0.3cm
\indent
\hskip 0.5cm
{\bf Acknowledgments}
\vskip 0.3cm
\indent
We are greatly indebted to Michel Dubois-Violette for numerous discussions and enlightening remarks.
We would like to thank Hubert Goenner and Dimitri V. Gal'tsov for pointing out the forgotten work
by R.R. Kuzeev. We thank Christian Klein for pertinent suggestions and constructive criticism.

\end{document}